\documentclass[preprint, aps, prd]{revtex4-1}

\def\pd{\partial}
\def\mc{\mathcal}

\usepackage[dvips]{graphicx}
\usepackage{amssymb}
\usepackage[bbgreekl]{mathbbol}
\usepackage{amssymb,amsmath}
\usepackage{graphicx}
\usepackage{dsfont}
\usepackage{caption}
\usepackage{subcaption}
\usepackage{mathtools}
\usepackage{verbatim}
\usepackage{graphicx}
\usepackage{multirow}
\usepackage[outline]{contour}
\usepackage{xcolor,colortbl}
\makeatother
\begin{document}

\title{Janus and RG-flow interfaces from 5D $N=4$ gauged supergravity}

\author{Parinya Karndumri} \email[REVTeX Support:
]{parinya.ka@hotmail.com} \affiliation{String Theory and
Supergravity Group, Department of Physics, Faculty of Science,
Chulalongkorn University, 254 Phayathai Road, Pathumwan, Bangkok
10330, Thailand}

\date{\today}
\begin{abstract}
We study five-dimensional $N=4$ gauged supergravity coupled to three vector multiplets with $SO(2)_D\times SO(3)$ gauge group. The gauged supergravity admits two supersymmetric $AdS_5$ vacua. One of the vacua, at the origin of the scalar manifold $\mathbb{R}^+\times SO(5,3)/SO(5)\times SO(3)$, preserves $N=4$ supersymmetry and the full $SO(2)_D\times SO(3)$ gauge symmetry. The second $AdS_5$ vacuum is $N=2$ supersymmetric and invariant under $SO(2)_{\textrm{diag}}$ which is a diagonal subgroup of $SO(2)_D$ and $SO(2)\subset SO(3)$. By considering a truncation to $SO(2)_{\textrm{diag}}$ invariant scalars, we find a class of Janus solutions interpolating among these vacua. These solutions should be holographically dual to conformal interfaces within the dual $N=1$ and $N=2$ SCFTs in four dimensions. In addition, we also find examples of solutions describing RG-flow interfaces between $N=1$ and $N=2$ SCFTs on each side of the interfaces. All of the solutions could possibly be uplifted to eleven dimensions by a consistent truncation of M-theory on $M_3\times S^2\times S^1$ with a 3-manifold $M_3$.
\end{abstract}
\maketitle
\section{Introduction}
The holographic study of conformal interfaces within strongly-coupled SCFTs is one of the most interesting aspects of the AdS/CFT correspondence \cite{maldacena,Gubser_AdS_CFT,Witten_AdS_CFT}. These interfaces correspond to a particular class of supergravity solutions called Janus solutions. In general, $(d-1)$-dimensional conformal interfaces in $d$-dimensional SCFTs can be obtained from $AdS_d$-sliced domain walls within a $(d+1)$-dimensional gauged supergravity. Since the discovery of the first Janus solution in \cite{Bak_Janus}, see also \cite{5D_Janus_CK}, many solutions of this type have appeared in various space-time dimensions, see \cite{Freedman_Janus}-\cite{Guarino_S_fold_Janus} for an incomplete list. Some of these solutions can be embedded in string/M-theory via many known consistent truncations. A number of Janus solutions have also been obtained directly from ten- or eleven-dimensional supergravities, see for example \cite{4D_Janus_from_11D,5D_Janus_DHoker1,5D_Janus_DHoker2,5D_Janus_DHoker3}. 
\\
\indent In this paper, we are interested in supersymmetric Janus solutions in half-maximal $N=4$ gauged supergravity in five dimensions. After the original supersymmetric Janus solution found in \cite{5D_Janus_CK}, see also \cite{5D_Janus_Suh}, a number of Janus solutions from the maximal $SO(6)$ gauged supergravity have appeared more recently in \cite{Bobev_5D_Janus2}. However, all of these solutions have been obtained within particular truncations of $N=2$ or $N=8$ gauged supergravities in which only one supersymmetric $AdS_5$ vacuum exists. Accordingly, the resulting Janus solutions essentially interpolate between the same conformal fixed point dual to the $AdS_5$ vacuum. Another class of solutions describing superconformal interfaces between $N=4$ Super-Yang-Mills and $N=1$ Leigh-Strassler SCFT have been found more recently in \cite{RG-interface_Gauntlett}. These solutions are also obtained from the $SO(6)$ maximal gauged supergravity in five dimensions. In this work, we will look for supersymmetric Janus solutions from $N=4$ gauged supergravity coupled to vector multiplets with more than one supersymmetric $AdS_5$ vacuum. These would provide first examples of Janus solutions involving more than one $AdS_5$ vacuum from the half-maximal $N=4$ gauged supergravity. 
\\
\indent We will consider $N=4$ gauged supergravity coupled to three vector multiplets with $SO(2)_D\times SO(3)\sim U(1)_D\times SU(2)$ gauge group. The full global symmetry of this matter-coupled supergravity is given by $\mathbb{R}^+\times SO(5,3)$. The gauge group under consideration here is embedded exclusively in the $SO(5,3)$ factor via it maximal compact subgroup $SO(5)\times SO(3)$. The $SO(5)$ factor corresponds to the R-symmetry $SO(5)_R\sim USp(4)_R$. The $SO(3)$ factor in the gauge group is identified with the $SO(3)_R$ subgroup of $SO(2)_R\times SO(3)_R\subset SO(5)_R$ while the $SO(2)_D$ is the diagonal subgroup of $SO(2)_R$ and $SO(2)\subset SO(3)$ with $SO(3)$ being the symmetry of the three vector multiplets. This gauged supergravity has originally been studied in \cite{5D_N4_flow_Davide} in which two supersymmetric $AdS_5$ vacua with $N=4$ and $N=2$ supersymmetries have been found. An extension to a bigger gauge group admitting a number of $N=4$ and $N=2$ supersymmetric $AdS_5$ vacua and a web of holographic RG flows has also been considered recently in \cite{5D_flowII}. In addition, it has been shown in \cite{Malek_AdS5_N4_embed} that this gauged supergravity can be obtained from a consistent truncation of eleven-dimensional supergravity on $M_3\times S^2\times S^1$ with $M_3$ being a three-manifold. It is then possible to uplift all the solutions found in this paper to M-theory.
\\
\indent The existence of the second $AdS_5$ vacuum leads to a much richer class of Janus solutions. Apart from the solutions interpolating between the $N=4$ $AdS_5$ vacuum at the origin of the scalar manifold $\mathbb{R}^+\times SO(5,3)/SO(5)\times SO(3)$ as in other previous works, there are solutions that approach arbitrarily close to the $N=2$ $AdS_5$ vacuum on both sides of the interfaces. This type of solutions has also been found in other dimensions, see for example \cite{warner_Janus,N3_JanusII,N4_omega_Janus}. In addition, we will find solutions describing conformal interfaces between different conformal fixed points related by RG flows. These solutions are called RG-flow interfaces and have attracted much attention recently \cite{6D_Janus_RG,3D_Janus2,3D_Janus3,3D_Janus4}. In addition to giving a new class of Janus solutions in $N=4$ gauged supergravity, we hope the results will give complements to the study of $AdS_5$ vacua and holographic RG flows as well as black string solutions given in \cite{5D_N4_flow_Davide,5D_flowII,5D_N4_black_stringII}. 
\\
\indent The paper is organized as follows. In section \ref{N4_SUGRA},
we review five-dimensional $N=4$ gauged supergravity coupled to three vector multiplets and $SO(2)_D\times SO(3)$ gauge group. The known $N=4$ and $N=2$ supersymmetric $AdS_5$ vacua are also reviewed. In section \ref{Janus_solutions}, we perform the analysis of supersymmetry conditions leading to a set of BPS equations for obtaining Janus solutions. We also give a number of numerical Janus solutions. We end the paper by giving some conclusions and comments in section \ref{conclusion}. 

\section{Five-dimensional $N=4$ gauged supergravity with $SO(2)_D\times SO(3)$ gauge group}\label{N4_SUGRA} 
We first give a brief review of five-dimensional $N=4$ gauged supergravity coupled to vector multiplets. We will closely follow the construction of $N=4$ gauged supergravity given in \cite{N4_gauged_SUGRA} and \cite{5D_N4_Dallagata} to which we refer for more details. There are only two types of multiplets in $N=4$ supersymmetry, the supergravity and vector multiplets. The former consists of the graviton $e^{\hat{\mu}}_\mu$, four gravitini $\psi_{\mu i}$, six vectors $(A_\mu^0,A_\mu^m)$, four spin-$\frac{1}{2}$ fields $\chi_i$ and one real scalar $\Sigma$, the dilaton. Each vector multiplet contains a vector field $A_\mu$, four gaugini $\lambda_i$ and five scalars $\phi^m$. The conventions on various types of indices are as follows. Space-time and tangent space indices are denoted respectively by $\mu,\nu,\ldots =0,1,2,3,4$ and
$\hat{\mu},\hat{\nu},\ldots=0,1,2,3,4$. The fundamental representation of $SO(5)_R\sim USp(4)_R$
R-symmetry is described respectively by $m,n=1,\ldots, 5$ for $SO(5)_R$ and $i,j=1,2,3,4$ for $USp(4)_R$. The vector multiplets are labeled by indices $a,b=1,\ldots, n$. 
\\
\indent In this paper, we are only interested in supersymmetric Janus solutions characterized exclusively by the metric and scalars. Therefore, we will set all the vector and tensor fields as well as spinor fields to zero. The dilaton can be described by a coorinate on $\mathbb{R}^+$ while the $5n$ scalar fields from the vector multiplets parametrize the $SO(5,n)/SO(5)\times SO(n)$ coset. The latter can be described by a coset representative $\mc{V}_M^{\phantom{M}A}$ transforming under the global $G=SO(5,n)$ and the local $H=SO(5)\times SO(n)$ by left and right multiplications, respectively. We use the global $SO(5,n)$ indices $M,N,\ldots=1,2,\ldots , 5+n$ while the local indices $A,B,\ldots$ can be split into $A=(m,a)$ as
\begin{equation}
\mc{V}_M^{\phantom{M}A}=(\mc{V}_M^{\phantom{M}m},\mc{V}_M^{\phantom{M}a}).
\end{equation}
\indent Gaugings of $N=4$ supergravity are described by three components of the embedding tensor $\xi^{M}$, $\xi^{MN}=\xi^{[MN]}$ and $f_{MNP}=f_{[MNP]}$. In order to define consistent gaugings, these components need to satisfy a set of quadratic constraints. In addition, the existence of supersymmetric $AdS_5$ vacua requires $\xi^M=0$ \cite{AdS5_N4_Jan}, so we will set $\xi^{M}=0$ from now on. The gauge group is then entirely embedded in $SO(5,n)$ with the quadratic constraints given by
\begin{equation}
f_{R[MN}{f_{PQ]}}^R=0\qquad \textrm{and}\qquad {\xi_M}^Qf_{QNP}=0\, .\label{QC}
\end{equation}
The gauge generators in the fundamental representation of $SO(5,n)$ can be written as
\begin{equation}
{(X_M)_N}^P=-{f_M}^{QR}{(t_{QR})_N}^P={f_{MN}}^P\quad \textrm{and}\quad {(X_0)_N}^P=-\xi^{QR}{(t_{QR})_N}^P={\xi_N}^P
\end{equation}
with ${(t_{MN})_P}^Q=\delta^Q_{[M}\eta_{N]P}$ being $SO(5,n)$ generators. We also note that the definition of $\xi^{MN}$ and $f_{MNP}$ includes the gauge coupling constants. $SO(5,n)$ indices $M,N,\ldots$ are lowered and raised by the invariant tensor $\eta_{MN}$ and its inverse $\eta^{MN}$ with $\eta_{MN}=\textrm{diag}(-1,-1,-1,-1,-1,1,\ldots,1)$
\\
\indent With vector and tensor fields vanishing, the bosonic Lagrangian of a general gauged $N=4$ supergravity coupled to vector multiplets can be written as
\begin{eqnarray}
e^{-1}\mc{L}&=&\frac{1}{2}R-\frac{3}{2}\Sigma^{-2}\pd_\mu \Sigma \pd^\mu \Sigma +\frac{1}{16} \pd_\mu M_{MN}\pd^\mu
M^{MN}-V\label{Lar}
\end{eqnarray}
with $e$ being the vielbein determinant and  
\begin{equation}
M_{MN}=\mc{V}_M^{\phantom{M}m}\mc{V}_N^{\phantom{M}m}+\mc{V}_M^{\phantom{M}a}\mc{V}_N^{\phantom{M}a}\, .
\end{equation}
The scalar potential reads
\begin{eqnarray}
V&=&\frac{1}{4}\left[f_{MNP}f_{QRS}\Sigma^{-2}\left(\frac{1}{12}M^{MQ}M^{NR}M^{PS}-\frac{1}{4}M^{MQ}\eta^{NR}\eta^{PS}
\right.\right.\nonumber \\ & &\left.
+\frac{1}{6}\eta^{MQ}\eta^{NR}\eta^{PS}\right) +\frac{1}{4}\xi_{MN}\xi_{PQ}\Sigma^4(M^{MP}M^{NQ}-\eta^{MP}\eta^{NQ})
\nonumber \\ & &\left.
+\frac{\sqrt{2}}{3}f_{MNP}\xi_{QR}\Sigma M^{MNPQR}\right]
\end{eqnarray}
where $M^{MN}$ is the inverse of $M_{MN}$. $M^{MNPQR}$ is defined as
\begin{equation}
M_{MNPQR}=\epsilon_{mnpqr}\mc{V}_{M}^{\phantom{M}m}\mc{V}_{N}^{\phantom{M}n}
\mc{V}_{P}^{\phantom{M}p}\mc{V}_{Q}^{\phantom{M}q}\mc{V}_{R}^{\phantom{M}r}
\end{equation}
with indices raised by $\eta^{MN}$. 
\\
\indent In order to obtain supersymmetric solutions from the corresponding BPS equations, we also need supersymmetry transformations of fermions which are given by
\begin{eqnarray}
\delta\psi_{\mu i} &=&D_\mu \epsilon_i+\frac{i}{\sqrt{6}}\Omega_{ij}A^{jk}_1\gamma_\mu\epsilon_k,\\
\delta \chi_i &=&-\frac{\sqrt{3}}{2}i\Sigma^{-1} \pd_\mu
\Sigma\gamma^\mu \epsilon_i+\sqrt{2}\Omega_{ij}A_2^{kj}\epsilon_k,\\
\delta \lambda^a_i&=&i\Omega^{jk}({\mc{V}_M}^a\pd_\mu
{\mc{V}_{ij}}^M)\gamma^\mu\epsilon_k+\sqrt{2}\Omega_{ij}A_{2}^{akj}\epsilon_k\,
.
\end{eqnarray}
The fermion shift matrices are defined by
\begin{eqnarray}
A_1^{ij}&=&-\frac{1}{\sqrt{6}}\left(\sqrt{2}\Sigma^2\Omega_{kl}{\mc{V}_M}^{ik}{\mc{V}_N}^{jl}\xi^{MN}+\frac{4}{3}\Sigma^{-1}{\mc{V}^{ik}}_M{\mc{V}^{jl}}_N{\mc{V}^P}_{kl}{f^{MN}}_P\right),\nonumber
\\
A_2^{ij}&=&\frac{1}{\sqrt{6}}\left(\sqrt{2}\Sigma^2\Omega_{kl}{\mc{V}_M}^{ik}{\mc{V}_N}^{jl}\xi^{MN}-\frac{2}{3}\Sigma^{-1}{\mc{V}^{ik}}_M{\mc{V}^{jl}}_N{\mc{V}^P}_{kl}{f^{MN}}_P\right),\nonumber
\\
A_2^{aij}&=&-\frac{1}{2}\left(\Sigma^2{{\mc{V}_M}^{ij}\mc{V}_N}^a\xi^{MN}-\sqrt{2}\Sigma^{-1}\Omega_{kl}{\mc{V}_M}^a{\mc{V}_N}^{ik}{\mc{V}_P}^{jl}f^{MNP}\right).
\end{eqnarray}
It is also useful to note that the scalar potential can be written in terms of these matrices as
\begin{equation}
V=-A^{ij}_1A_{1ij}+A^{ij}_2A_{2ij}+A_2^{aij}
A^a_{2ij}\, .
\end{equation}
\indent The coset representative of the form $\mc{V}_M^{\phantom{M}ij}$ is defined in terms of ${\mc{V}_M}^m$ and $SO(5)$ gamma matrices ${\Gamma_{mi}}^j$ as
\begin{equation}
{\mc{V}_M}^{ij}=\frac{1}{2}{\mc{V}_M}^{m}\Gamma^{ij}_m
\end{equation}
with $\Gamma^{ij}_m=\Omega^{ik}{\Gamma_{mk}}^j$. Similarly, the inverse ${\mc{V}_{ij}}^M$ can be written as
\begin{equation}
{\mc{V}_{ij}}^M=\frac{1}{2}{\mc{V}_m}^M(\Gamma^{ij}_m)^*=\frac{1}{2}{\mc{V}_m}^M\Gamma_{m}^{kl}\Omega_{ki}\Omega_{lj}\,
.
\end{equation}
As in \cite{5D_flowII}, we use the following representation of ${\Gamma_{mi}}^j$ matrices
\begin{eqnarray}
\Gamma_1&=&-\sigma_2\otimes \sigma_2,\qquad \Gamma_2=\mathbb{I}_2\otimes \sigma_1,\qquad \Gamma_3=\mathbb{I}_2\otimes \sigma_3,\nonumber\\
\Gamma_4&=&\sigma_1\otimes \sigma_2,\qquad \Gamma_5=\sigma_3\otimes \sigma_2
\end{eqnarray}
with $\sigma_i$, $i=1,2,3$, being the Pauli matrices. It is also useful to note that raising and lowering of $i,j,\ldots$ indices by $\Omega^{ij}$ and $\Omega_{ij}$ are related to complex conjugation. We also use an explicit form of $\Omega_{ij}$ given by
\begin{equation}
\Omega_{ij}=\Omega^{ij}=i\sigma_2\otimes \mathbb{I}_2\, .
\end{equation}
\indent The covariant derivative on $\epsilon_i$ is given by
\begin{equation}
D_\mu \epsilon_i=\pd_\mu \epsilon_i+\frac{1}{4}\omega_\mu^{ab}\gamma_{ab}\epsilon_i+{Q_{\mu i}}^j\epsilon_j
\end{equation}
with the composite connection defined by
\begin{equation}
{Q_{\mu i}}^j={\mc{V}_{ik}}^M\pd_\mu {\mc{V}_M}^{kj}\, .
\end{equation}
\indent We end this section by reviewing the $N=4$ gauged supergravity coupled to three vector multiplets with $SO(2)_D\times SO(3)$ gauge group. In the notation of \cite{5D_flowII}, the embedding tensor is given by
\begin{eqnarray}
\xi^{MN}&=&-g_1(\delta^M_1\delta^N_2-\delta^M_2\delta^N_1)+g_2(\delta^M_{6}\delta^N_{7}-\delta^M_{7}\delta^N_{6}),\\ 
f_{\tilde{m}+2,\tilde{n}+2,\tilde{p}+2}&=&-h_1\epsilon_{\tilde{m}\tilde{n}\tilde{p}},\qquad \tilde{m},\tilde{n},\tilde{p}=1,2,3
\end{eqnarray} 
with $g_1$, $g_2$ and $h_1$ being coupling constants. The $SO(3)$ factor is identified with an $SO(3)_R$ subgroup of $SO(2)_R\times SO(3)_R\subset SO(5)_R$ while the $SO(2)_D$ is a diagonal subgroup of the $SO(2)_R$ and the $SO(2)$ subgroup of the $SO(3)$ symmetry of the three vector multiplets. Within the framework of five-dimensional $N=4$ gauged supergravity, the three coupling constants are independent. However, upon uplifting to eleven dimensions, the coupling constants $g_1$ and $h_1$ are fixed in terms of the radius of the $N=4$ $AdS_5$ vacuum at the origin of the scalar manifold. On the other hand, the coupling constant $g_2$ is related to the $U(1)_D\sim SO(2)_D$ charge of the first two vector multplets charged under $SO(2)_D$, see \cite{Malek_AdS5_N4_embed} for more detail. It is useful to note that, in the present convention, for $g_2=-g_1$, the resulting gauged supergravity can be embedded in the maximal $SO(6)$ gauged supergravity as pointed out in \cite{5D_N4_flow_Davide}. In this case, the corresponding solutions can also be embedded in type IIB theory via the well-known consistent truncation on $S^5$ to the maximal gauged supergravity. 
\\
\indent To give a concrete parametrization of the $SO(5,3)/SO(5)\times SO(3)$ coset, we write the $SO(5,3)$ non-compact generators as
\begin{equation}
Y_{ma}=t_{m,a+5},\qquad m=1,2,\ldots, 5,\qquad a=1,2,3\, .
\end{equation}
By choosing the parametrization, as considered in \cite{5D_N4_flow_Davide},
\begin{equation}
\mc{V}=e^{\phi (Y_{41}+Y_{52})},\label{truncated_coset}
\end{equation}  
we find the scalar potential
\begin{eqnarray}
V&=&\frac{1}{4}\frac{\cosh^2\phi}{\Sigma^2}\left[(\cosh2\phi-3)h_1^2+4\sqrt{2}g_1h_1\Sigma^3+2g_2^2\Sigma^6\sinh^2\phi\right]
\end{eqnarray}
which admits two supersymmetric $AdS_5$ vacuum given by
\begin{eqnarray}
\phi=0,\qquad \Sigma=1,\qquad V_0=-3g_1^2,\qquad L=\frac{\sqrt{2}}{g_1}\label{N4_AdS5}
\end{eqnarray}
and
\begin{eqnarray}
& &\phi=\frac{1}{2}\ln\left[\frac{g_2-4g_1+2\sqrt{4g_1^2-2g_1g_2-2g_2^2}}{3g_2}\right],\qquad \Sigma=-\left(\frac{2g_1}{g_2}\right)^{\frac{1}{3}},\nonumber \\
& &V_0=-\frac{1}{3}(g_1-g_2)^2\left(\frac{2g_1}{g_2}\right)^{\frac{4}{3}},\qquad L=\frac{3}{g_1-g_2}\left(\frac{g_2^2}{\sqrt{2}g_1^2}\right)^{\frac{1}{3}}\label{N2_AdS5}
\end{eqnarray}
The first vacuum preserves the full $N=4$ supersymmetry and $SO(2)_D\times SO(3)$ gauge symmetry. The second vacuum is on the other hand $N=2$ supersymmetric and breaks the gauge group to $SO(2)_{\textrm{diag}}\subset [SO(2)_D\times SO(2)]_{\textrm{diag}}$ with $SO(2)\subset SO(3)$. 
\\
\indent In the above equations, we have also set $h_1=-\sqrt{2}g_1$ in order to set $\Sigma=1$ at the $N=4$ supersymmetric $AdS_5$ vacuum, see more detail in \cite{5D_flowII}. We have also used the relation between the cosmological constant $V_0$ and the $AdS_5$ radius $L$ given by
\begin{equation}
L=\sqrt{-\frac{6}{V_0}}
\end{equation}
with a definite choice of $g_1>g_2$. These two vacua should be dual to $N=2$ and $N=1$ SCFTs in four dimensions, respectively. Some aspects of the possible dual SCFTs have been given in \cite{5D_N4_flow_Davide}. Moreover, holographic RG flows between these two vacuum have already been studied in \cite{5D_N4_flow_Davide} and \cite{5D_flowII}.
\section{Supersymmetric Janus solutions}\label{Janus_solutions}
In this section, we look for supersymmetric Janus solutions describing conformal interfaces within the $N=2$ and $N=1$ SCFTs dual to the aforementioned $AdS_5$ vacua. The metric ansatz is given by an $AdS_4$-sliced domain wall
\begin{equation} 
ds^2=e^{2A(r)}ds^2_{AdS_4}+dr^2
\end{equation}
with $ds^2_{AdS_4}$ being the metric on $AdS_4$ with radius $\ell$. 
\\
\indent We will consider a truncation to $SO(2)_{\textrm{diag}}$ singlet scalars. This diagonal subgroup is generated by a linear combination of $SO(5,3)$ generators $t_{12}+t_{45}+t_{67}$ under which there are five singlet scalars from $SO(5,3)/SO(5)\times SO(3)$ coset. This can be seen as follows. We recall that the $15$ scalars transform as $(\mathbf{5},\mathbf{3})$ under the compact subgroup $SO(5)\times SO(3)\subset SO(5,3)$. Decomposing $\mathbf{5}$ of $SO(5)\sim SO(5)_R$ under the $SO(2)_R\times SO(3)_R$ subgroup via $\mathbf{5}\rightarrow (\mathbf{2},\mathbf{1})+(\mathbf{1},\mathbf{3})$, we find that
\begin{equation}
(\mathbf{5},\mathbf{3})\rightarrow (\mathbf{2},\mathbf{1},\mathbf{3})+(\mathbf{1},\mathbf{3},\mathbf{3}).
\end{equation} 
Further decomposing $SO(3)_R$ and $SO(3)$ to $SO(2)'_R$ and $SO(2)$ subgroups via $\mathbf{3}\rightarrow \mathbf{1}+\mathbf{2}$, we end up with the transformation of all these $15$ scalars under $SO(2)_R\times SO(2)'_R\times SO(2)$ 
\begin{equation}
(\mathbf{5},\mathbf{3})\rightarrow (\mathbf{2},\mathbf{1},\mathbf{1})+(\mathbf{2},\mathbf{1},\mathbf{2})+(\mathbf{1},\mathbf{1},\mathbf{1})+(\mathbf{1},\mathbf{2},\mathbf{2})+(\mathbf{1},\mathbf{2},\mathbf{1})+(\mathbf{1},\mathbf{1},\mathbf{2}).
\end{equation} 
We now form the diagonal subgroup of all the $SO(2)_R$, $SO(2)'_R$ and $SO(2)$ factors by taking a tensor product of all three representations in each factor. Together with the fact that the product of $SO(2)$ fundamental representation $\mathbf{2}\otimes \mathbf{2}\rightarrow \mathbf{1}+\mathbf{1}+\mathbf{2}$, we eventually find five singlets under $SO(2)_{\textrm{diag}}$ coming from $(\mathbf{1},\mathbf{1},\mathbf{1})$, $(\mathbf{2},\mathbf{1},\mathbf{2})$ and $(\mathbf{1},\mathbf{2},\mathbf{2})$. These singlets correspond to the following non-compact generators
\begin{eqnarray}
& &\hat{Y}_1=Y_{11}+Y_{22},\qquad \hat{Y}_2=Y_{41}+Y_{52},\qquad \hat{Y}_3=Y_{33},\nonumber \\
& &\hat{Y}_4=Y_{12}-Y_{21},\qquad \hat{Y}_5=Y_{42}-Y_{51}\, . 
\end{eqnarray}
The coset representative can then be written as
\begin{equation}
\mc{V}=e^{\phi_1\hat{Y}_1}e^{\phi_2\hat{Y}_2}e^{\phi_3\hat{Y}_3}e^{\phi_4\hat{Y}_4}e^{\phi_5\hat{Y}_5}\, .\label{coset_rep}
\end{equation}
It could be useful to note that the five scalars parametrize an $SO(1,1)\times SU(2,1)/SU(2)\times U(1)$ submanifold of $SO(5,3)/SO(5)\times SO(3)$. The generator $\hat{Y}_3$ corresponds to the $SO(1,1)$ factor while the remaining four generators $\hat{Y}_i$ for $i=1,2,4,5$ that commute with $\hat{Y}_3$ correspond to non-compact generators of $SU(2,1)$. The $SU(2)\times U(1)$ compact generators, on the other hand, are given by $(t_{14}+t_{25},t_{15}-t_{24},t_{12}-t_{45})$ and $t_{12}+t_{45}-2t_{67}$, respectively. 
\\
\indent It turns out that the presence of $\phi_4$ and $\phi_5$ scalars leads to non-vanishing Yang-Mills currents. Therefore, in order to consistently truncate out all the gauge fields, we need to set $\phi_4=\phi_5=0$. With $\phi_4=\phi_5=0$, a direct computation leads to the scalar potential
\begin{eqnarray}
V&=&\frac{1}{8}\Sigma^{-1}\left[h_1^2\cosh2\phi_3\sinh^22\phi_2+\cosh^2\phi_2\left[-4h_1^2\right.\right.\nonumber \\
& &+4\sqrt{2}h_1(g_1+g_2+(g_1-g_2)\cosh2\phi_1)\cosh\phi_3\Sigma^3\nonumber \\
& &+\left\{g_2^2\cosh^2\phi_1(\cosh2\phi_1-3)+2g_2^2\cosh^4\phi_1\cosh2\phi_2\right.\nonumber \\
& &+g_1^2\sinh^2\phi_1(3+\cosh2\phi_1+2\cosh2\phi_2\sinh^2\phi_1)\nonumber \\ 
& &\left.\left.\left. -2g_1g_2\cosh^2\phi_2\sinh^22\phi_1 \right\}\Sigma^6\right]\right].\label{poten}
\end{eqnarray}
The coset representative \eqref{truncated_coset} is a subtruncation of the one given in \eqref{coset_rep}, so the scalar potential \eqref{poten} admits the same $AdS_5$ critical points as given in \eqref{N4_AdS5} and \eqref{N2_AdS5}. Apart from these two critical points, there are no other supersymmetric $AdS_5$ vacua.
\\
\indent It is also useful to note the scalar kinetic term
\begin{equation}
\mc{L}_{\textrm{kin}}=-\frac{3}{2}\frac{\Sigma'^2}{\Sigma^2}-\cosh^2\phi_2{\phi'_1}^2-{\phi'_2}^2-\frac{1}{2}{\phi'_3}^2\, .
\end{equation}

\subsection{BPS equations for Janus solutions}
We are now in a position to analyze the supersymmetry conditions arising from setting supersymmetry transformations of fermionic fields to zero. With the coset representative \eqref{coset_rep} and $\phi_4=\phi_5=0$, the $A_1^{ij}$ tensor is diagonal and takes the form
\begin{equation}
A^{ij}_1=\textrm{diag}(\alpha,\beta,\alpha^*,\beta)\label{A1_tensor}
\end{equation}
with
\begin{eqnarray}
\alpha&=&\frac{1}{\sqrt{6}}h_1\Sigma^{-1}\cosh^2\phi_2\cosh\phi_3+\frac{1}{8\sqrt{3}}\Sigma^2\left[(g_1-g_2)\cosh2\phi_1(\cosh2\phi_2-3)\right.\nonumber \\
& &\left.-2(g_1+g_2)\cosh^2\phi_2-4i(g_1-g_2)\sinh2\phi_1\sinh\phi_2 \right],\\
\beta&=&\frac{1}{8\sqrt{3}}\Sigma^2\left[2(g_1-g_2)\cosh2\phi_1\cosh^2\phi_2-(g_1+g_2)(\cosh2\phi_2-3) \right]\nonumber \\
& &-\frac{1}{\sqrt{6}}\Sigma^{-1}h_1\cosh^2\phi_2\cosh\phi_3\, .
\end{eqnarray}
As pointed out in \cite{5D_flowII}, the unbroken supersymmetry is at most $N=2$ whenever $\phi_2\neq 0$. 
\\
\indent However, unlike the previous results given in \cite{5D_N4_flow_Davide} and \cite{5D_flowII}, both the eigenvalues $\alpha$ and $\beta$ do not lead to a superpotential in terms of which the scalar potential \eqref{poten} can be written. It is possible to define an exact superpotential from $\alpha$ or $\beta$ only for either $\phi_1=0$ or $\phi_2=0$. As we will see, these two possibilities rule out any possible supersymmetric Janus solutions. Therefore, we will keep both $\phi_1$ and $\phi_2$ non-vanishing. The non-existence of superpotential is mainly due to the failure to satisfy some integrability conditions as pointed out in \cite{N2_holography_Bobev}. This is also the case for Janus solutions in six-dimensional gauged supergravity studied in \cite{6D_Janus} and \cite{6D_Janus_RG} as pointed out in \cite{mass_deform_5D_SCFT}. In addition, the reality of $\beta$ turns out to exclude the possibility of any Janus solutions as well. Accordingly, we will only consider the unbroken supersymmetry corresponding to the eigevalue $\alpha$ and take the associated Killing spinors to be $\epsilon_1$ and $\epsilon_3$.   
\\
\indent With all these, the variations $\delta\chi_i$ give 
\begin{equation}
\Sigma'\gamma_{\hat{r}}\epsilon_1=\mc{A}\epsilon_3\qquad \textrm{and}\qquad \Sigma'\gamma_{\hat{r}}\epsilon_3=\mc{A}^*\epsilon_1\label{Sigma_eq1}
\end{equation}
for 
\begin{eqnarray}
\mc{A}&=&-\frac{i}{6\sqrt{2}}\Sigma^3\left[g_2(2i\sinh\phi_1+2\cosh\phi_1\sinh\phi_2)^2 \right.\nonumber \\
& &\left.+4g_1(\cosh\phi_1-i\sinh\phi_1\sinh\phi_2)^2 \right]-\frac{i}{3}h_1\cosh^2\phi_2\cosh\phi_3\, .
\end{eqnarray}
Multiply the first equation in \eqref{Sigma_eq1} by $\Sigma'\gamma_{\hat{r}}$ and use the second equation, we find 
\begin{equation}
{\Sigma'}^2=\mc{A}\mc{A}^*=|\mc{A}|^2
\end{equation}
which leads to
\begin{equation}
\Sigma'=\eta |\mc{A}|
\end{equation}  
for a sign factor $\eta=\pm 1$ and a projector of the form
\begin{equation}
\gamma_{\hat{r}}\epsilon_1=\eta \frac{\mc{A}}{|\mc{A}|}\epsilon_3\qquad \textrm{and}\qquad \gamma_{\hat{r}}\epsilon_3=\eta \frac{\mc{A}^*}{|\mc{A}|}\epsilon_1\, .\label{gamma_r_proj}
\end{equation}
\indent We now move to the conditions arising from $\delta\lambda^a_i$. Using the $\gamma_{\hat{r}}$ projection given in \eqref{gamma_r_proj}, we find that one of the conditions leads to
\begin{equation}
(g_2-g_1)\sinh2\phi_1\sinh\phi_2\phi'_3=0.
\end{equation}
For $g_2=g_1$, the $N=2$ supersymmetric $AdS_5$ vacuum does not exist. As we will see at the end of this section, Janus solutions do not exist for $g_2=g_1$ either. Furthermore, Janus solutions do not exist if either $\phi_1=0$ or $\phi_2=0$ as previously mentioned. Therefore, we satisfy this condition by choosing constant $\phi_3$ which can be consistently set to zero in order to make the solutions approach the $AdS_5$ vacua at which $\phi_3=0$.  
\\
\indent The remaining conditions from $\delta \lambda^a_i$ are given by
\begin{eqnarray}
& & (\phi'_2-i\cosh\phi_2\phi'_1)\gamma_{\hat{r}}\epsilon_1=\mc{B}\epsilon_3\\
\textrm{and}\qquad & & (\phi'_2+i\cosh\phi_2\phi'_1)\gamma_{\hat{r}}\epsilon_3=\mc{B}^*\epsilon_1
\end{eqnarray}
with
\begin{eqnarray}
\mc{B}&=&\frac{1}{4\sqrt{2}}\Sigma^2\cosh\phi_2\left[2i\sinh\phi_2[(g_1-g_2)\cosh2\phi_1-g_1-g_2] \right.\nonumber \\
& &\left. -2(g_1-g_2)\sinh2\phi_1\right]+\frac{i}{2}\Sigma^{-1}h_1\sinh2\phi_2\, .
\end{eqnarray}
Using the projector \eqref{gamma_r_proj} again, we arrive at the BPS equations for $\phi_1$ and $\phi_2$ of the form
\begin{eqnarray}
\phi'_2\Sigma'=\textrm{Re}(\mc{B}\mc{A}^*)\qquad \textrm{and}\qquad \cosh\phi_2\phi'_1\Sigma'=-\textrm{Im}(\mc{B}\mc{A}^*).
\end{eqnarray}
\indent Finally, we consider the gravitino variations along $AdS_4$ directions. We first split the five-dimensional coordinates as $x^\mu=(x^\alpha,r)$ for $\alpha=0,1,2,3$. The conditions $\delta\psi_{\hat{\alpha}i}$ take the form
\begin{equation}
D_{\hat{\alpha}}\epsilon_i=-\frac{i}{\sqrt{6}}\Omega_{ij}A_1^{jk}\gamma_{\hat{\alpha}}\epsilon_k\, .\label{gravi_eq1}
\end{equation}
Following \cite{Bobev_5D_Janus2}, we rewrite the covariant derivative in terms of the covariant derivative $\widetilde{\nabla}_\alpha$ on $AdS_4$ as 
\begin{equation}
D_\alpha\epsilon_i=\widetilde{\nabla}_\alpha\epsilon_i-\frac{1}{2}A'\gamma_{r}\gamma_{\alpha}\epsilon_i\, .
\end{equation}
We then use the Killing spinor equations for $AdS_4$ of the form
\begin{equation}
\widetilde{\nabla}_\alpha\epsilon_i=\frac{i}{2\ell}\kappa_i\gamma_{r}\gamma_{\alpha}\epsilon_i
\end{equation}
with $\kappa_i=\pm 1$. We have also identified the chirality matrix on $AdS_4$ as $\gamma_r=i\gamma_{\hat{0}}\gamma_{\hat{1}}\gamma_{\hat{2}}\gamma_{\hat{3}}$. Using the explicit form of the $A_1^{ij}$ tensor given in \eqref{A1_tensor}, we find from \eqref{gravi_eq1}
\begin{eqnarray}
& &\left(A'-\frac{i}{\ell}\kappa_1 e^{-A}\right)\gamma_{\hat{r}}\epsilon_1=-i\mc{W}^*\epsilon_3\label{gravitino_eq1}\\
\textrm{and}\qquad & & \left(A'-\frac{i}{\ell}\kappa_3 e^{-A}\right)\gamma_{\hat{r}}\epsilon_3=i\mc{W}\epsilon_1\label{gravitino_eq2}
\end{eqnarray}
with $\mc{W}=\sqrt{\frac{2}{3}}\alpha$. As in \cite{Bobev_5D_Janus2}, consistency between these two equations requires $\kappa_3=-\kappa_1$. 
\\
\indent Using the $\gamma_{\hat{r}}$ projector given in \eqref{gamma_r_proj} and writing $\kappa=\kappa_1=-\kappa_3$, we eventually find
\begin{equation}
A'=-\eta \frac{\textrm{Re}(i\mc{W}^*\mc{A}^*)}{|\mc{A}|}\qquad \textrm{and}\qquad \frac{\kappa}{\ell}e^{-A}=\eta\frac{\textrm{Im}(i\mc{W}^*\mc{A}^*)}{|\mc{A}|}\, .\label{BPS_gravi}
\end{equation} 
We also note that the two equations \eqref{gravitino_eq1} and \eqref{gravitino_eq2} also imply the relation
\begin{equation}
{A'}^2+\frac{1}{\ell^2}e^{-2A}=|\mc{W}|^2\, .
\end{equation}
As in other cases, the remaining condition $\delta\psi_{\hat{r}i}$ will determine the radial dependence of the Killing spinors.
\\
\indent At this point, we will give some comments on choosing the residual supersymmetry corresponding to the eigenvalue $\beta$ of the $A_1^{ij}$ tensor. In this case, the Killing spinors are given by $\epsilon_2$ and $\epsilon_4$. The same analysis of $\delta \chi_i$ conditions leads to an analogue of \eqref{Sigma_eq1} of the form 
\begin{equation}
\Sigma'\gamma_{\hat{r}}\epsilon_2=i\mc{M}\epsilon_4\qquad \textrm{and}\qquad \Sigma'\gamma_{\hat{r}}\epsilon_4=-i\mc{M}\epsilon_2 
\end{equation}
with a real function $\mc{M}$. Consistency of the projectors then requires that $\Sigma'=\pm \mc{M}$ which gives the projectors 
\begin{equation}
\gamma_{\hat{r}}\epsilon_2=\pm i\epsilon_4\qquad \textrm{and}\qquad \gamma_{\hat{r}}\epsilon_4=\mp i\epsilon_2 \, .
\end{equation}
Using these projectors in $\delta \psi_{\hat{\alpha}i}$ conditions results in the identically vanishing $\frac{\kappa}{\ell}e^{-A}$. Therefore, only flat domain wall solutions are possible.  
\\
\indent It is useful to explicitly give the form of the algebraic constraint given by the second equation in \eqref{BPS_gravi}
\begin{equation}
\frac{\kappa}{\ell}e^{-A}=\frac{(g_2-g_1)h_1\Sigma^2\sinh2\phi_1\sinh\phi_2\cosh^2\phi_2}{3\sqrt{2}|\mc{A}|}\, .\label{constraint}
\end{equation}
We can immediately see that for either $\phi_1=0$ or $\phi_2=0$, this constraint forces the $AdS_4$ radius $\ell\rightarrow \infty$ leading to a flat domain wall. This also implies that the dilaton cannot support the curvature of the world-volume of the domain wall. This is precisely in agreement with the results of \cite{5D_N4_curved_DW} in which it has also been shown that the existence of $AdS_4$-sliced domain walls requires both the dilaton and scalars from vector multiplets to be non-vanishing. Furthermore, setting $g_2=g_1$ also excludes any Janus solutions as previously mentioned.  
\\
\indent We end this section by pointing out that the algebraic constraint \eqref{constraint} is compatible with all the remaining flow equations for $\Sigma$, $\phi_1$, $\phi_2$ and $A$. Moreover, we have also explicitly verified that these equations together with the constraint \eqref{constraint} imply the second-ordered field equations. Finally, it should be noted that Janus solutions are also possible for $g_2=0$. This leads to a simple embedding of the $SO(2)$ factor, and the $SO(2)\times SO(3)$ gauge group is entirely embedded in $SO(5)_R\subset SO(5,3)$. Therefore, supersymmetric Janus solutions exist in both $SO(2)\times SO(3)$ and $SO(2)_D\times SO(3)$ gauge groups provided that both $\phi_1$ and $\phi_2$ are non-vanishing. 
\subsection{Numerical Janus solutions}
\begin{figure}
         \centering
         \begin{subfigure}[b]{0.35\textwidth}
                 \includegraphics[width=\textwidth]{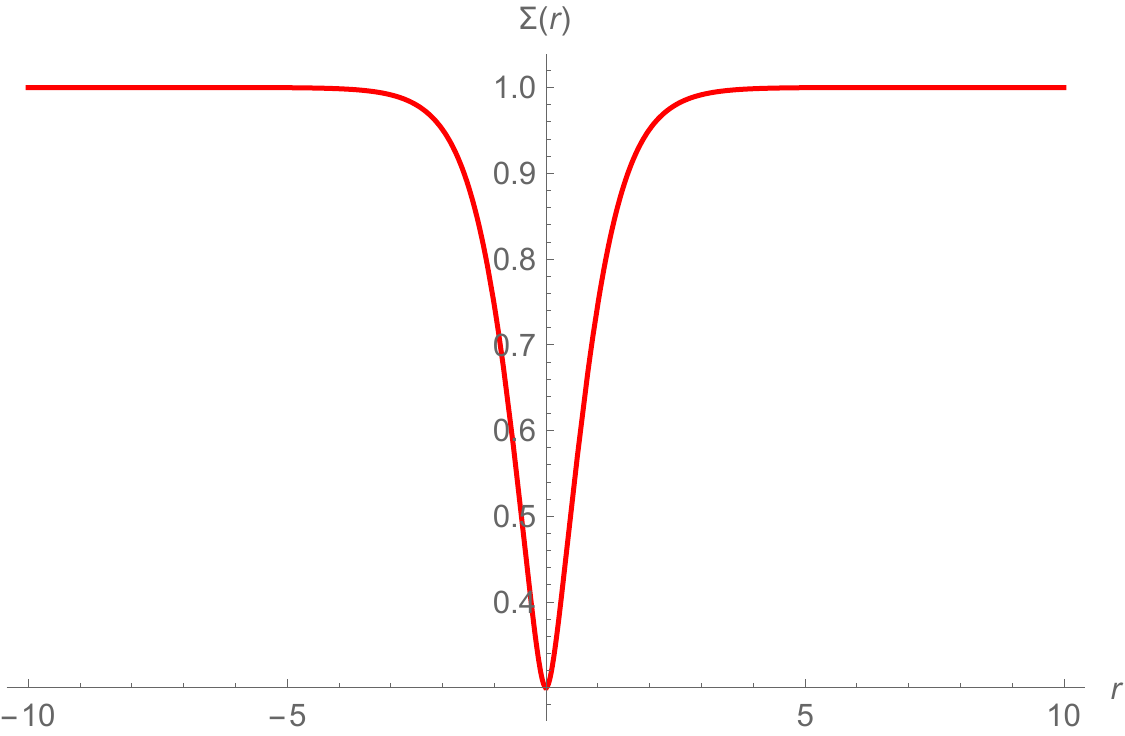}
                 \caption{Solution for $\Sigma(r)$}
         \end{subfigure} \qquad
\begin{subfigure}[b]{0.35\textwidth}
                 \includegraphics[width=\textwidth]{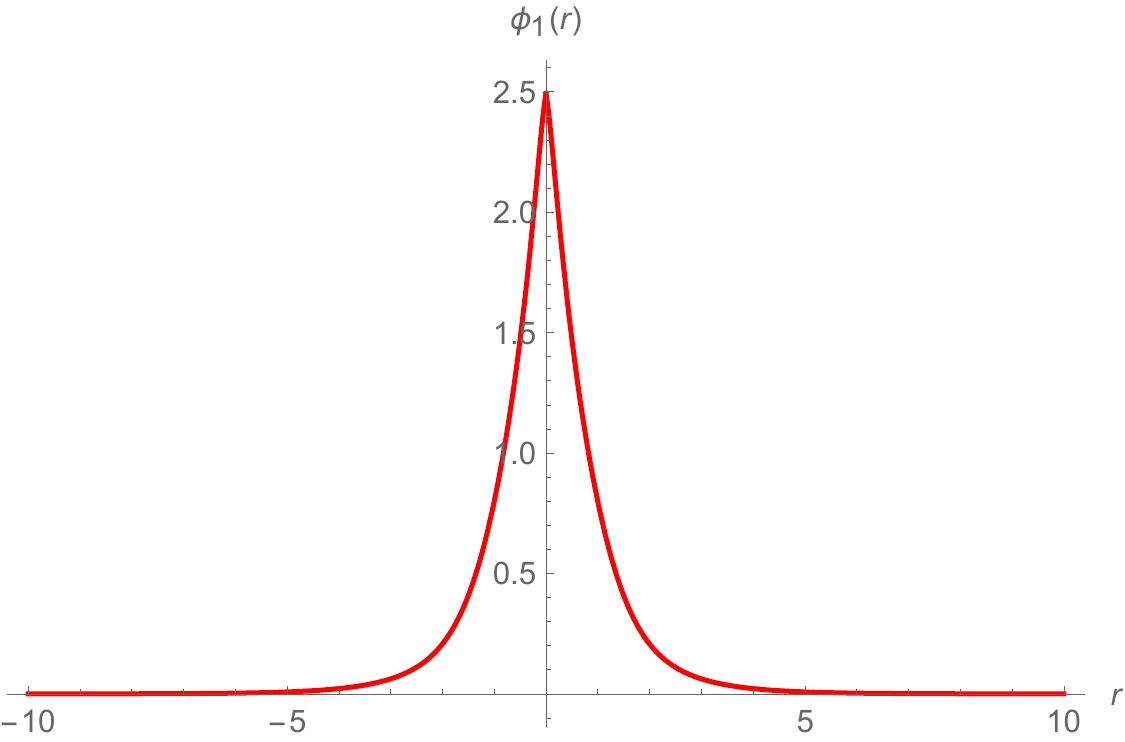}
                 \caption{Solution for $\phi_1(r)$}
         \end{subfigure}\\
         \begin{subfigure}[b]{0.35\textwidth}
                 \includegraphics[width=\textwidth]{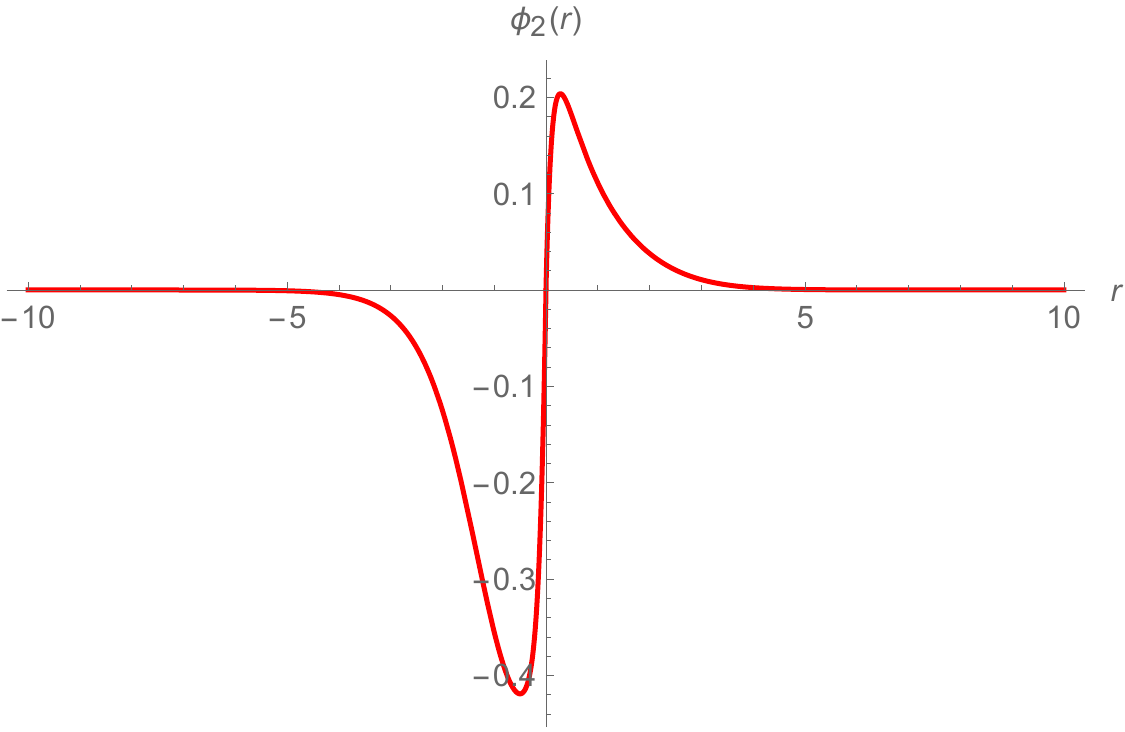}
                 \caption{Solution for $\phi_2(r)$}
         \end{subfigure}\qquad 
         \begin{subfigure}[b]{0.35\textwidth}
                 \includegraphics[width=\textwidth]{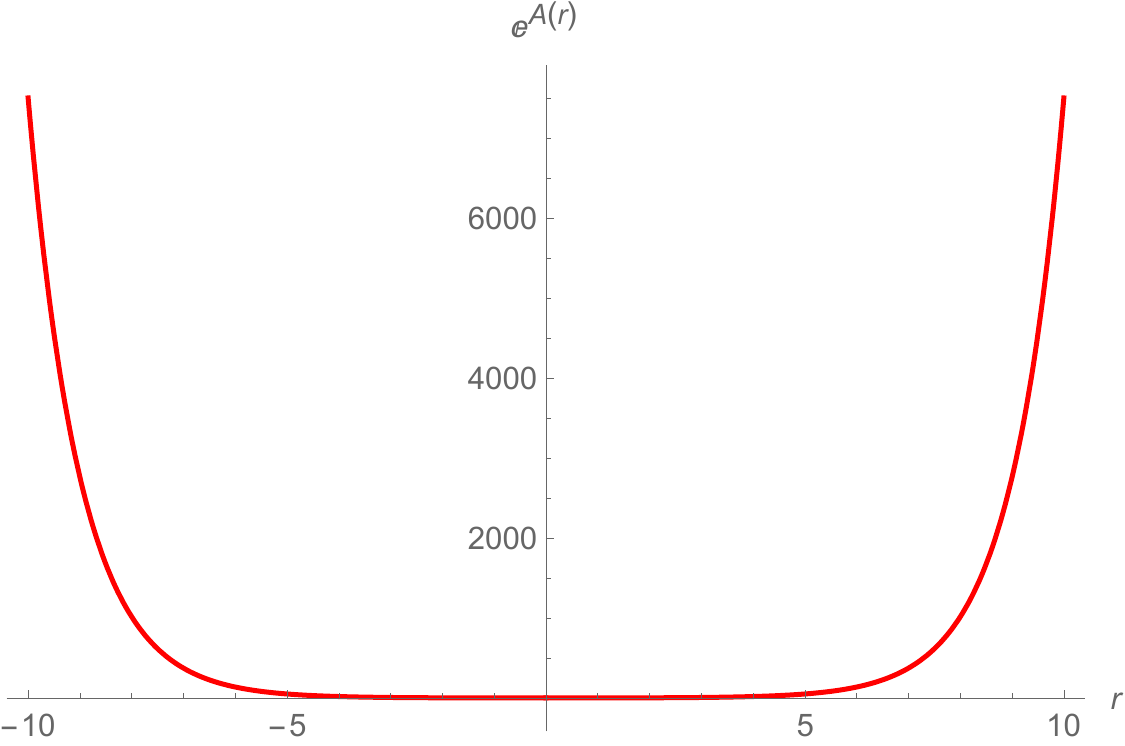}
                 \caption{Solution for $e^{A(r)}$}
         \end{subfigure} 
         \caption{An example of Janus solutions interpolating between $N=4$ $AdS_5$ vacua in $SO(2)\times SO(3)$ gauge group with $\ell=\kappa=1$, $h_1=-2$, $g_1=\sqrt{2}$ and $g_2=0$.}\label{fig1}
 \end{figure}

\begin{figure}
         \centering
         \begin{subfigure}[b]{0.35\textwidth}
                 \includegraphics[width=\textwidth]{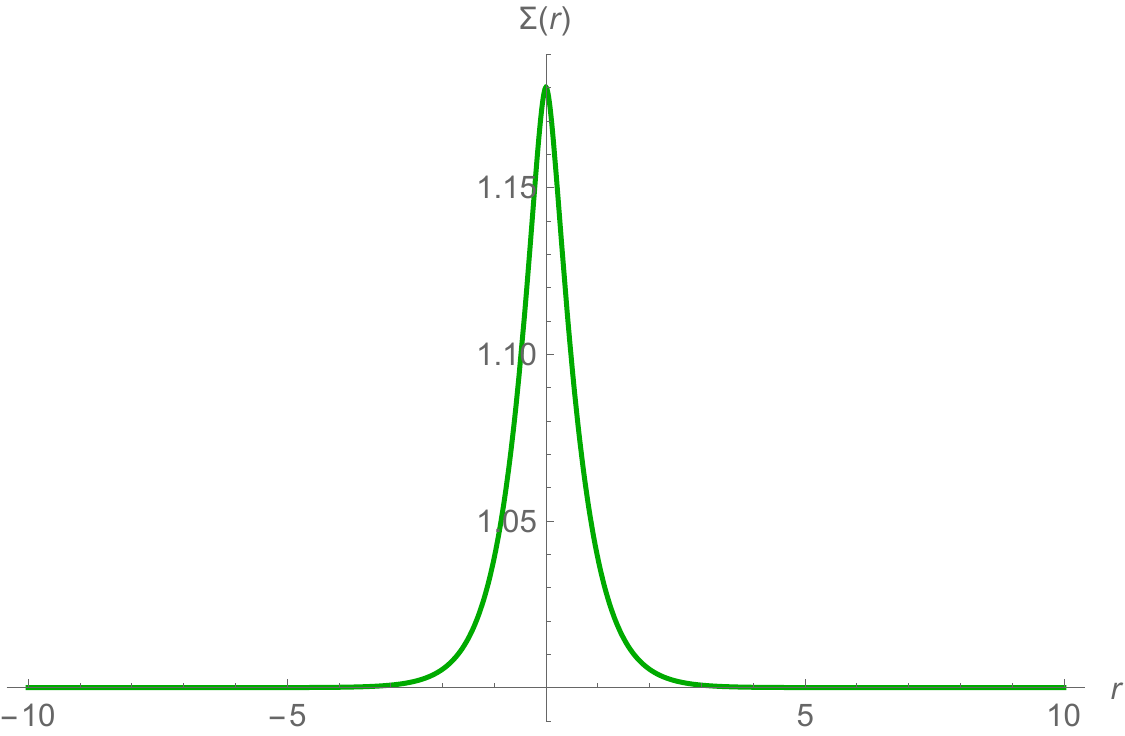}
                 \caption{Solution for $\Sigma(r)$}
         \end{subfigure} \qquad
\begin{subfigure}[b]{0.35\textwidth}
                 \includegraphics[width=\textwidth]{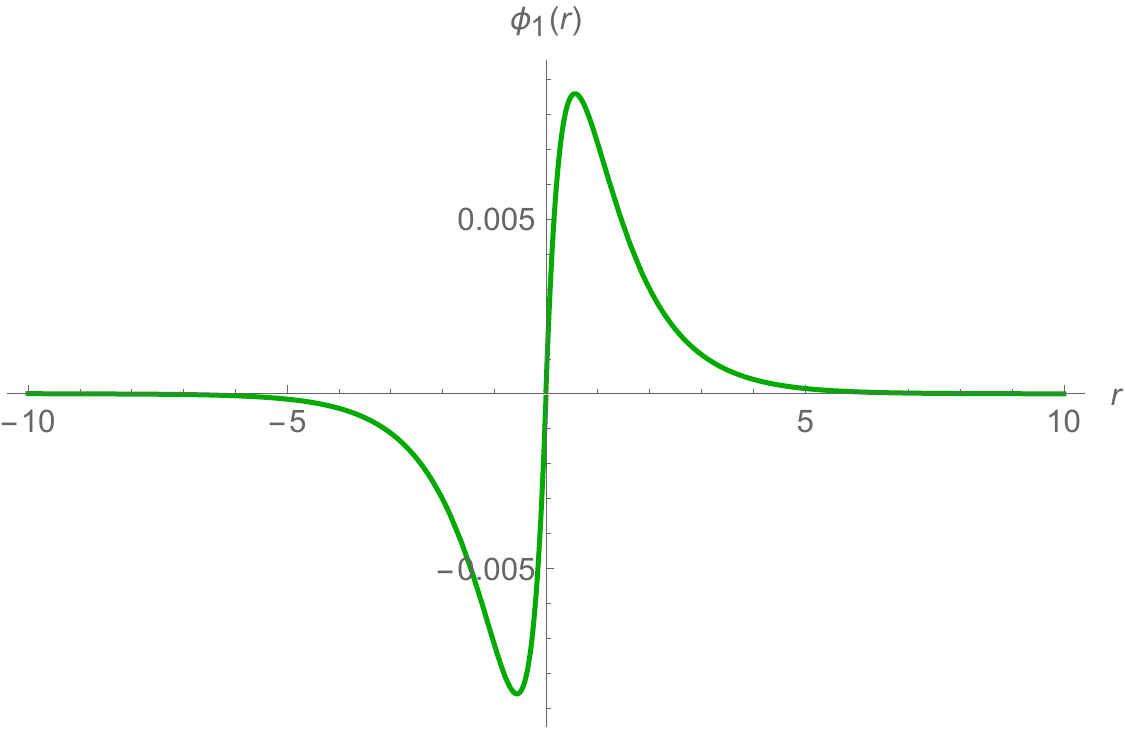}
                 \caption{Solution for $\phi_1(r)$}
         \end{subfigure}\\
         \begin{subfigure}[b]{0.35\textwidth}
                 \includegraphics[width=\textwidth]{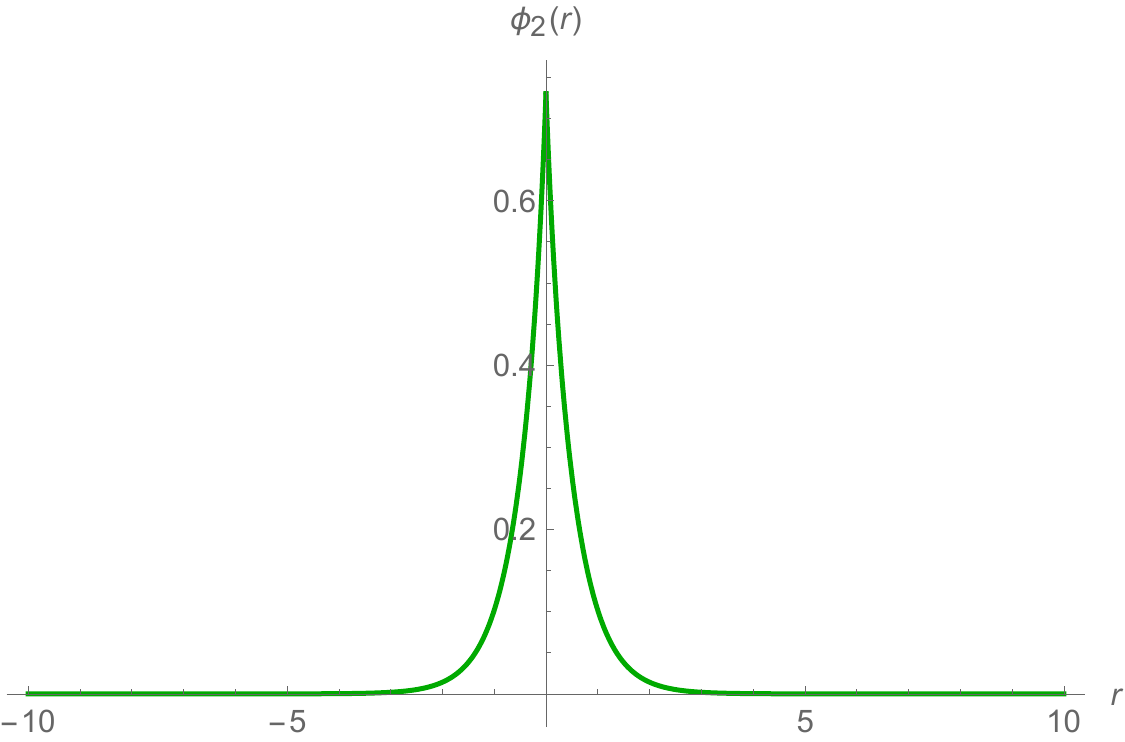}
                 \caption{Solution for $\phi_2(r)$}
         \end{subfigure}\qquad 
         \begin{subfigure}[b]{0.35\textwidth}
                 \includegraphics[width=\textwidth]{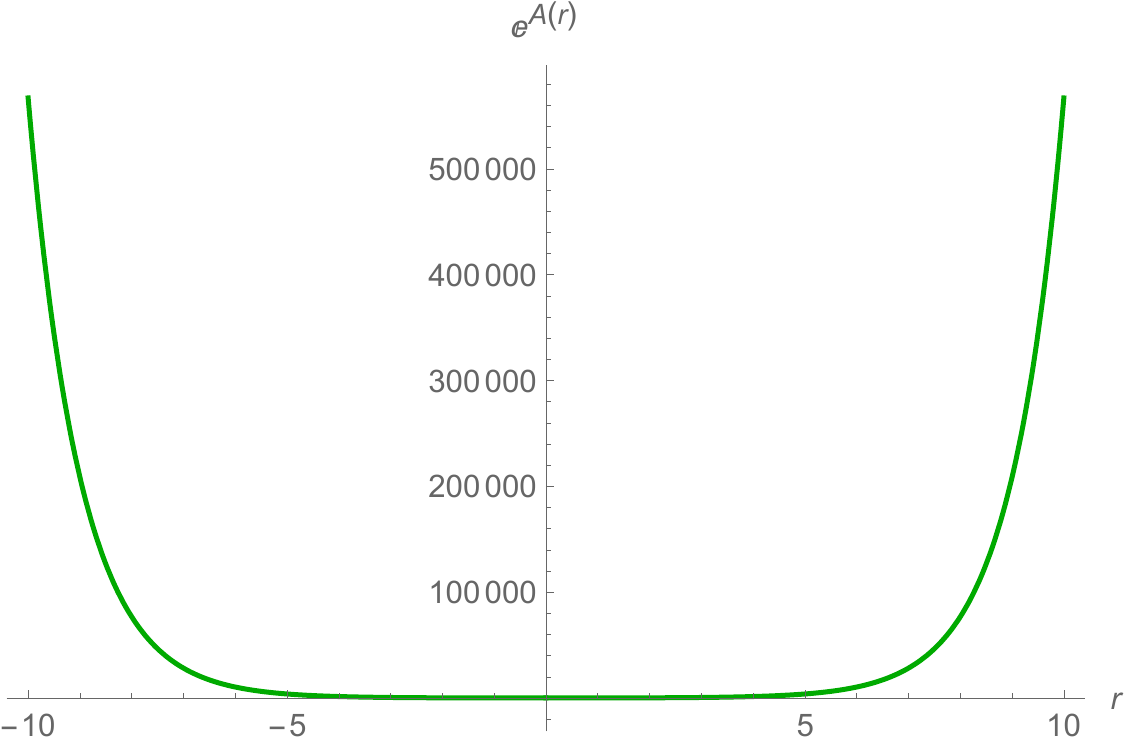}
                 \caption{Solution for $e^{A(r)}$}
         \end{subfigure} 
         \caption{Another example of Janus solutions interpolating between $N=4$ $AdS_5$ vacua in $SO(2)\times SO(3)$ gauge group with $\ell=\kappa=1$, $h_1=-2$, $g_1=\sqrt{2}$ and $g_2=0$.}\label{fig2}
 \end{figure}

The resulting BPS equations take a very complicated form. Even a much simpler set of BPS equations for obtaining RG flow solutions studied in \cite{5D_N4_flow_Davide} can only be solved numerically. We then do not expect to find any analytic solutions to these equations. Accordingly, we consider finding numerical solutions with suitable boundary conditions. For convenience, we first note the explicit form of all the BPS equations 
\begin{eqnarray}
\Sigma'&=& \frac{1}{3}\eta\left[\left(\sqrt{2}\Sigma^3\{\cosh^2\phi_1(g_1+g_2\sinh^2\phi_2)-\sinh^2\phi_1(g_2+g_1\sinh^2\phi_2)\}\right.\right.\nonumber \\
& &\left.\left. \phantom{\sqrt{2}}+h_1\cosh^2\phi_2\right)^2 +2(g_1-g_2)^2\Sigma^6\sinh^22\phi_1\sinh^2\phi_2\right]^{\frac{1}{2}},\\
\Sigma'\phi'_1&=&\frac{1}{24}(g_1-g_2)\frac{\Sigma^2\cosh\phi_2\sinh2\phi_1}{\cosh\phi_2}\left[\sqrt{2}h_1(5\cosh2\phi_2-3) \right.\nonumber \\
& &\left.\phantom{\sqrt{2}}+\Sigma^3[2(g_1-g_2)\cosh2\phi_1\cosh^2\phi_2-(g_1+g_2)(\cosh2\phi_2-3)] \right],\\
\Sigma'\phi'_2&=&\frac{1}{24\sqrt{2}}h_1\sinh2\phi_2\Sigma^2\left[(g_1-g_2)\cosh2\phi_1(\cosh2\phi_2-7)\right.\nonumber \\
& &\left.-2(g_1+g_2)\cosh^2\phi_2\right]+\frac{1}{96}\Sigma^5\left[2\sinh2\phi_2(3g_1-g_2\right.\nonumber \\
& &+(g_1-g_2)\cosh2\phi_1)(3g_2-g_1+(g_1-g_2)\cosh2\phi_1)\nonumber \\
& &\left. +(g_1+g_2+(g_2-g_1)\cosh2\phi_1)^2\sinh4\phi_2\right]-\frac{h_1^2}{3\Sigma}\cosh^3\phi_2\sinh\phi_2,\\
\Sigma'A'&=&\frac{h_1^2}{9\Sigma}\cosh^4\phi_2+\frac{1}{72\sqrt{2}}h_1\Sigma^2\cosh^2\phi_2\left[4(g_1+g_2)\cosh^2\phi_2\right. \nonumber \\
& &\left.-2(g_1-g_2)\cosh2\phi_1(\cosh2\phi_2-3)\right]-\frac{1}{144}\Sigma^5\left[16(g_1-g_2)^2\sinh^22\phi_1\times\right.\nonumber \\
 & &\left.\times \sinh^2\phi_2+((g_1-g_2)\cosh2\phi_1(\cosh2\phi_2-3)-2(g_1+g_2)\cosh^2\phi_2)^2 \right]\nonumber \\
 & &
\end{eqnarray}
together with an algebraic constraint
\begin{equation}
\frac{\kappa}{\ell}e^{-A}=\frac{(g_2-g_1)h_1\Sigma^2\cosh^2\phi_2\sinh2\phi_1\sinh\phi_2}{3\sqrt{2}|\Sigma'|}
\end{equation}
for $\eta=\pm 1$ and $\kappa=\pm 1$. It is also useful to note that the solutions preserve $\frac{1}{4}$ of the original supersymmetry or four supercharges due to the projector \eqref{gamma_r_proj} and the fact that only $\epsilon_1$ and $\epsilon_3$ lead to the Killing spinors. In addition, with $\phi_1$ and $\phi_2$ non-vanishing, the solutions break $SO(2)_D\times SO(3)$ gauge symmetry to $SO(2)_{\textrm{diag}}$ subgroup. Therefore, the solutions holographically describe three-dimensional $N=1$ conformal interfaces within $N=1$ or $N=2$ SCFTs in four dimensions. 
\\
\indent We will numerically solve these equations by specifying the values of all the fields at a turning point with $A'(r_0)=0$. Without loss of generality, we can also choose $r_0=0$. As in other cases, most of the boundary conditions will lead to singular solutions in which the warp factor $A(r)$ or scalars diverge at some finite value of $r$. These solutions might holographically describe boundary SCFTs, but in this paper, we will only consider regular Janus solutions which approach $AdS_5$ vacua on both sides of the interfaces. The $AdS_5$ vacua on the two sides could be the same or different. In the latter, the resulting solutions are dual to RG-flow interfaces which interpolate between two conformal fixed points related by an RG flow. 
\\
\indent To obtain regular Janus solutions, we need to smoothly sew the two branches of solutions corresponding to $\eta=\pm1$ together. Alternatively, we can avoid this branch cut by solving second-ordered field equations with the boundary conditions for $\phi'_1(0)$, $\phi'_2(0)$ and $\Sigma'(0)$ obtained from the BPS equations as pointed out in \cite{warner_Janus}. Since the BPS equations take a similar form to those considered in \cite{6D_Janus} and \cite{6D_Janus_RG}, we will follow the same procedure for finding numerical solutions. We begin with solutions in the case of $g_2=0$ corresponding to $SO(2)\times SO(3)$ gauge group. As previously mentioned, in this case, the $N=2$ supersymmetric $AdS_5$ vacuum does not exist. Accordingly, we can have only solutions interpolating between $N=4$ supersymmetric $AdS_5$ vacuum on each side of the interfaces. For definiteness, we will set the $AdS_4$ radius $\ell=1$ and $\kappa=1$ without loss of any generality. Recall that we also have $h_1=-\sqrt{2}g_1$. We will further choose $g_1=\sqrt{2}$ in order to set the radius of the $N=4$ $AdS_5$ vacuum to unity. As in \cite{6D_Janus} and \cite{6D_Janus_RG}, it turns out that only some particular values of $\phi_1(0)$, $\phi_2(0)$, $\Sigma(0)$ and $A(0)$ lead to regular Janus solutions. With $g_2=0$, a numerical search for regular solutions leads to examples of Janus solutions as shown in figures \ref{fig1} and \ref{fig2}. These solutions should correspond to conformal interfaces in $N=2$ SCFT in four dimensions.   
\\
\indent Near the $N=4$ $AdS_5$ vacuum, the BPS equations give asymptotic behaviors of scalar fields as
\begin{eqnarray}
& &\Sigma \sim -\frac{1}{3}e^{-\frac{2r}{L}}\left(\frac{2}{L}C_1^2r-C_\Sigma\right),\qquad L=\frac{\sqrt{2}}{g_1},\nonumber \\
& &\phi_1\sim C_1e^{-\frac{r}{L}},\qquad \phi_2\sim C_2e^{-\frac{2r}{L}}\label{N4_asymp}
\end{eqnarray}
with $C_\Sigma$ and $C_{1,2}$ being constants. We see that the deformations involve turning on a dimension-$2$ operator dual to $\Sigma$ and a dimension-$3$ operator dual to $\phi_1$. The vacuum expectation values of the operators dual to both $\Sigma$ and $\phi_2$ are also present. We also see that all source terms vanish if $\phi_1=0$ ($C_1=0$).

\begin{figure}
         \centering
               \begin{subfigure}[b]{0.45\textwidth}
                 \includegraphics[width=\textwidth]{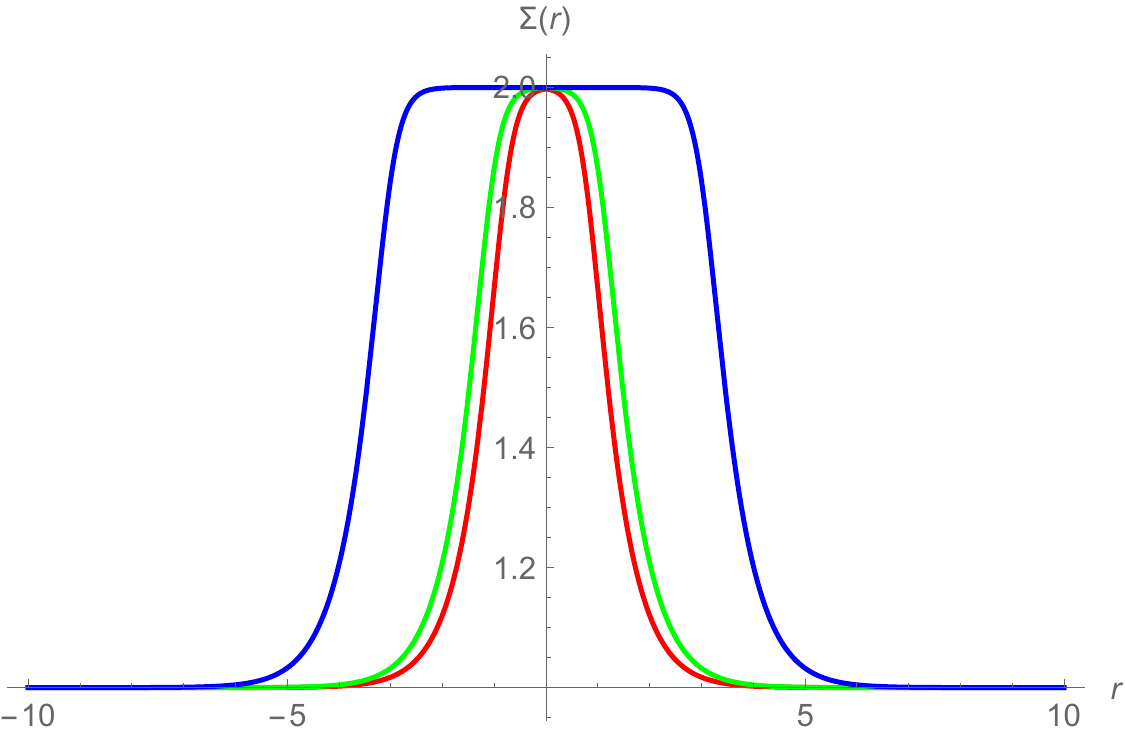}
                 \caption{Solutions for $\Sigma(r)$}
         \end{subfigure}
         \begin{subfigure}[b]{0.45\textwidth}
                 \includegraphics[width=\textwidth]{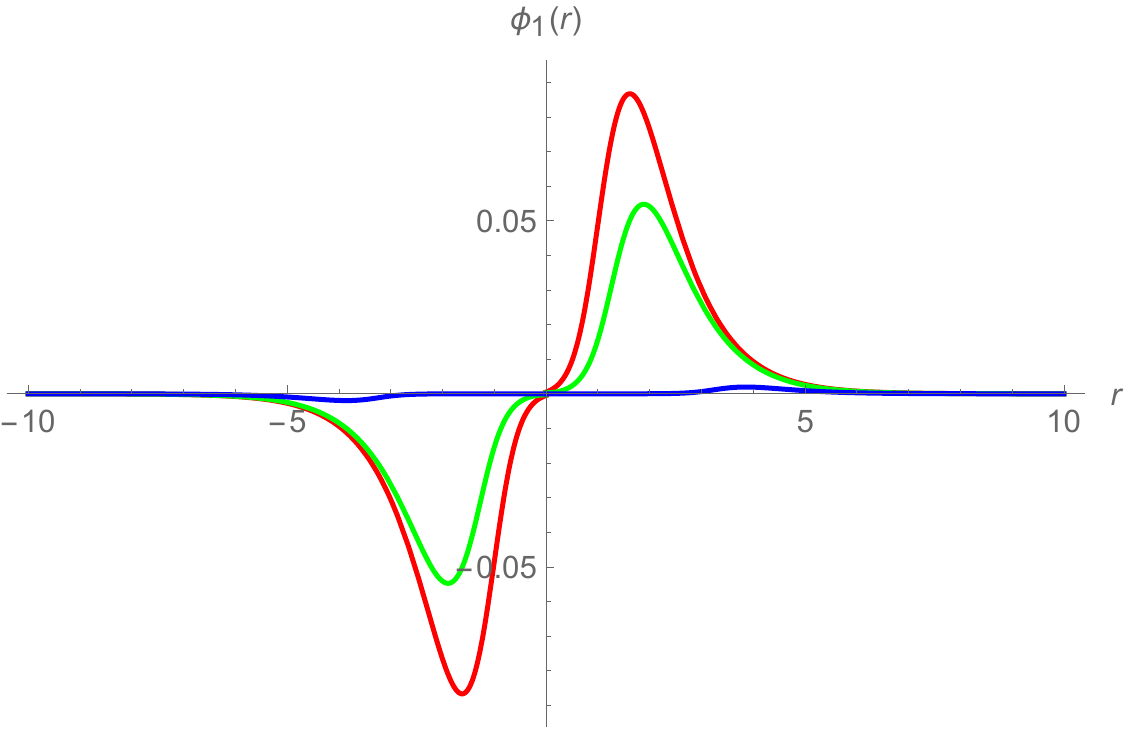}
                 \caption{Solutions for $\phi_1(r)$}
         \end{subfigure}\\
         \begin{subfigure}[b]{0.45\textwidth}
                 \includegraphics[width=\textwidth]{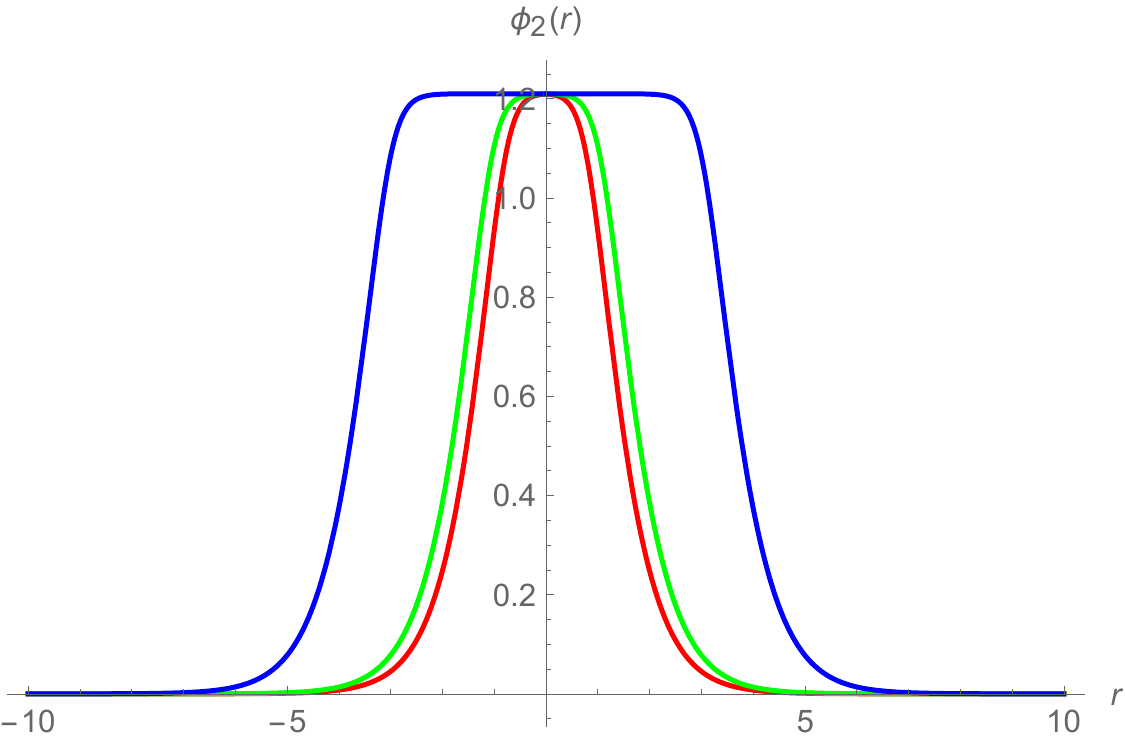}
                 \caption{Solutions for $\phi_2(r)$}
         \end{subfigure}
         \begin{subfigure}[b]{0.45\textwidth}
                 \includegraphics[width=\textwidth]{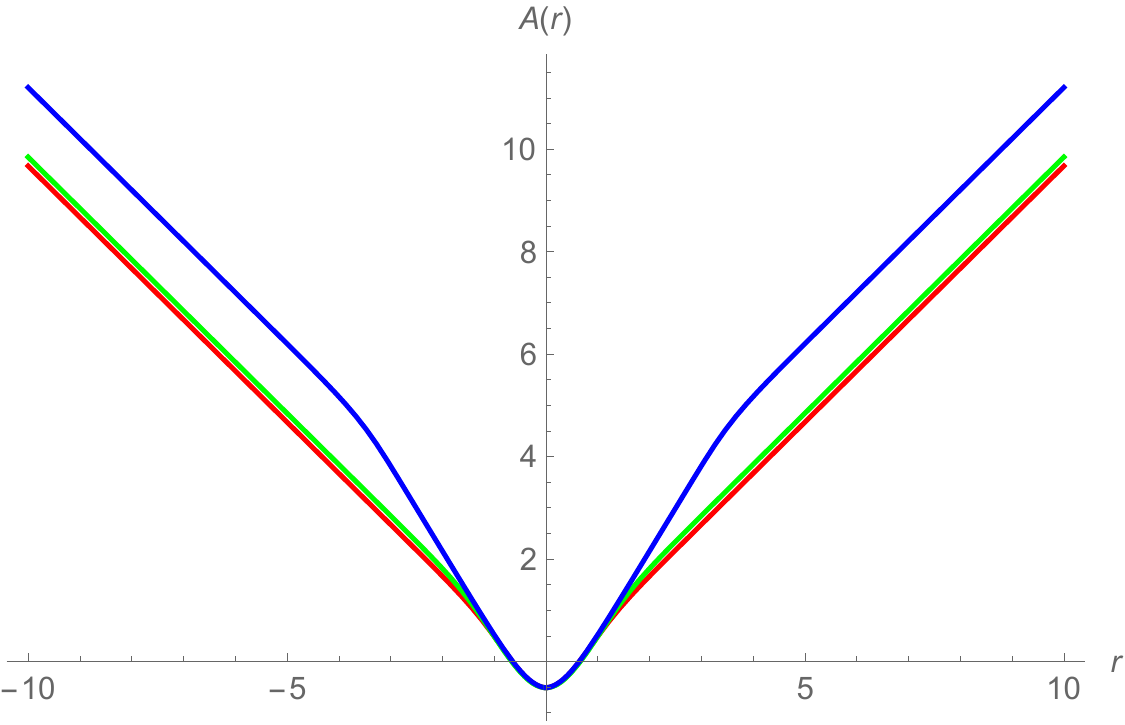}
                 \caption{Solutions for $A(r)$}
         \end{subfigure}\\
          \begin{subfigure}[b]{0.45\textwidth}
                 \includegraphics[width=\textwidth]{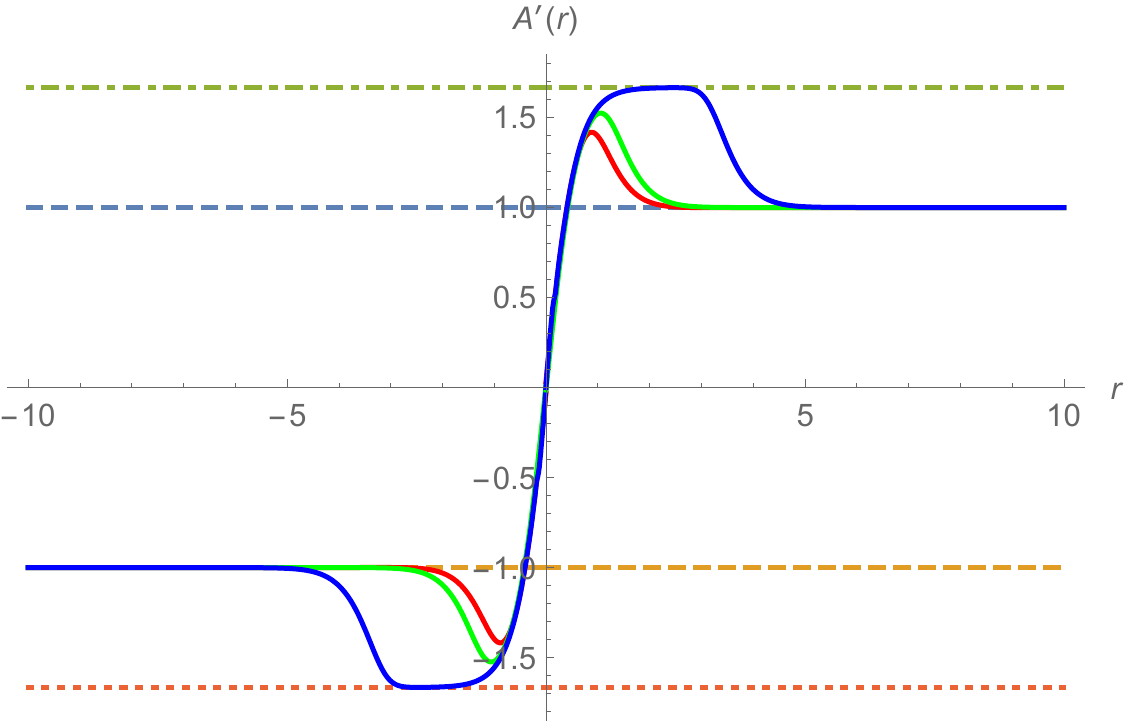}
                 \caption{Solutions for $A'(r)$}
         \end{subfigure}
         \caption{Examples of supersymmetric Janus solutions interpolating between $N=4$ $AdS_5$ vacua and approaching $N=2$ $AdS_5$ vacuum near the interfaces in $SO(2)_D\times SO(3)$ gauge group with $\ell=\kappa=1$, $h_1=-2$, $g_1=\sqrt{2}$ and $g_2=-\frac{\sqrt{2}}{4}$.}\label{fig3}
 \end{figure}

\indent We now consider another class of solutions for $SO(2)_D\times SO(3)$ gauge group with $g_2\neq 0$. In this case, there are two supersymmetric $AdS_5$ vacua with $N=4$ and $N=2$ supersymmetries. We then expect more interesting possible solutions. Unlike the other parameters, the coupling constant $g_2$ is a free parameter provided that its value still gives rise to a valid $N=2$ $AdS_5$ vacuum. The numerical analysis indicates that the solutions with different values of $g_2$ are not qualitatively different. Therefore, we will choose a particular values of $g_2=-\frac{1}{4}g_1$ with the other parameters being the same as in the previous case with $g_2=0$. One reason for this choice is that it leads to the values of $AdS_5$ radii at the $N=4$ and $N=2$ vacua sufficiently different from each other. As a consequence, it is more convenient to distinguish between solutions interpolating among different $AdS_5$ vacua. We also recall that for $g_2=-g_1$ the $N=4$ gauged supergravity can be embedded in the maximal $SO(6)$ gauged supergravity as previously mentioned. In this case, the solutions should correspond to Janus and RG-flow interfaces studied in \cite{RG-interface_Gauntlett} and can also be uplifted to solutions of type IIB string theory. For other values of $g_2$ considered in this work, the resulting solutions can only be embedded in eleven dimensions. 
\\
\indent Due to the presence of the $N=2$ $AdS_5$ vacuum, apart from the solutions interpolating directly between the $N=4$ $AdS_5$ vacua as in the previous case, there are solutions interpolating between $N=4$ $AdS_5$ vacua that flow very close to the $N=2$ $AdS_5$ vacuum. Examples of these solutions are shown in figure \ref{fig3}. When the position of the turning point is closer to the $N=2$ $AdS_5$ critical point, the solutions approach the $N=2$ critical point closer and stay near the $N=2$ critical point longer as can be seen respectively from the red, green and blue curves. As in similar solutions in other dimensions, we interpret these solutions as conformal interfaces between $N=2$ $AdS_5$ vacua on each side of the interfaces with the $N=2$ phases on both sides being generated by standard holographic RG flows from the $N=4$ phases. The horizontal dashed lines represent the values of $\frac{1}{L}$ at the two $AdS_5$ critical points.  

\begin{figure}
         \centering
               \begin{subfigure}[b]{0.35\textwidth}
                 \includegraphics[width=\textwidth]{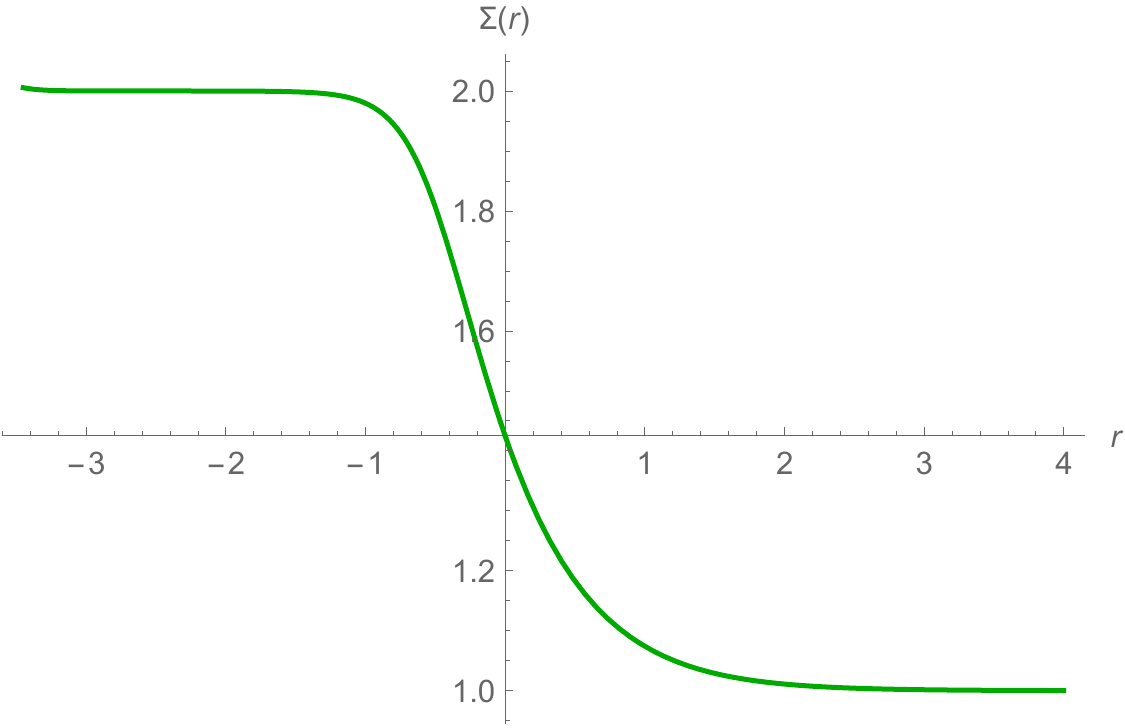}
                 \caption{Solution for $\Sigma(r)$}
         \end{subfigure}
         \begin{subfigure}[b]{0.35\textwidth}
                 \includegraphics[width=\textwidth]{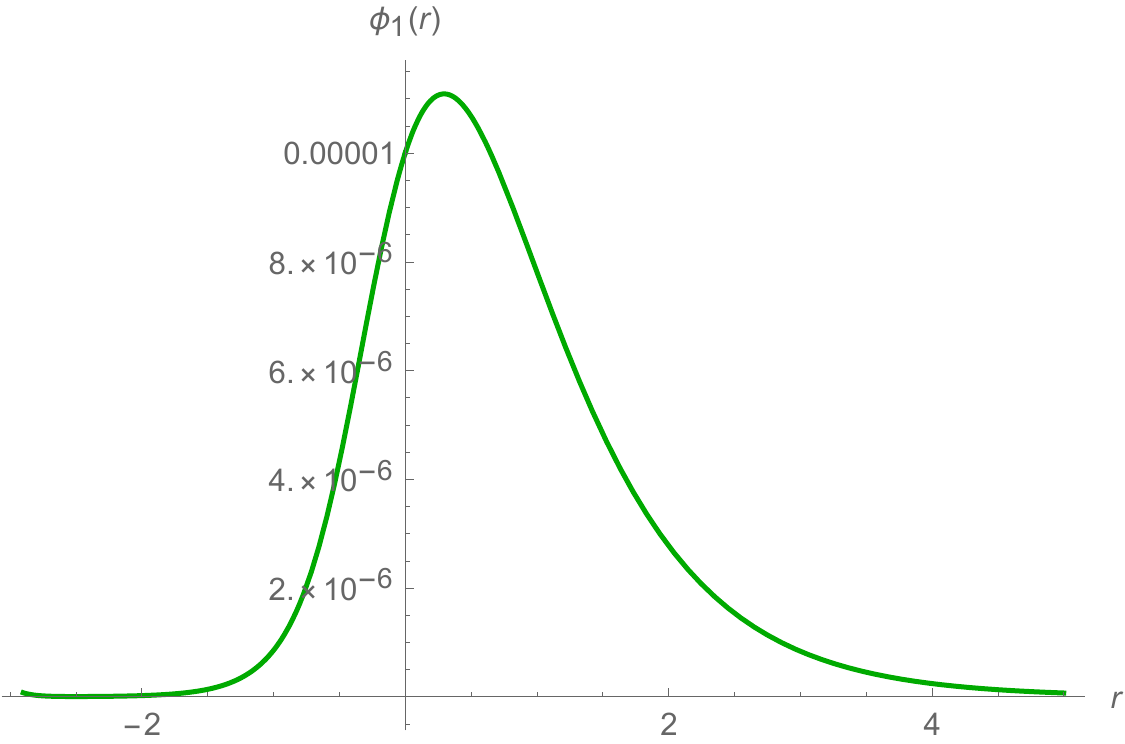}
                 \caption{Solution for $\phi_1(r)$}
         \end{subfigure}\\
         \begin{subfigure}[b]{0.35\textwidth}
                 \includegraphics[width=\textwidth]{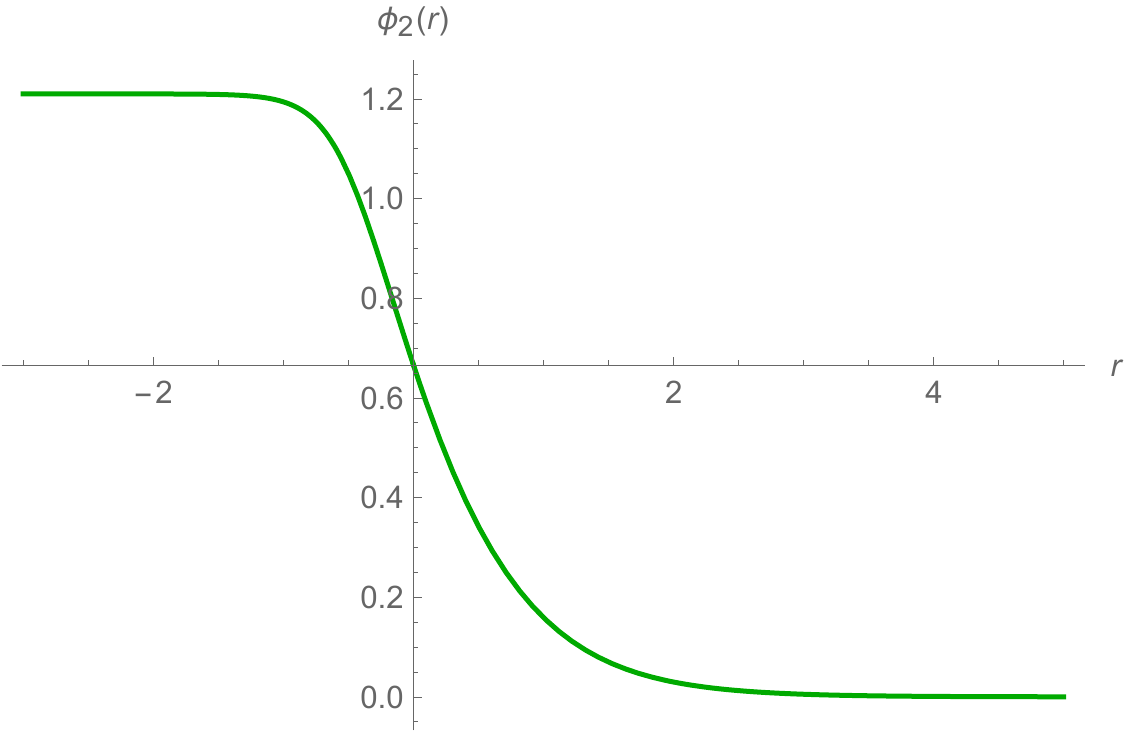}
                 \caption{Solution for $\phi_2(r)$}
         \end{subfigure}
         \begin{subfigure}[b]{0.35\textwidth}
                 \includegraphics[width=\textwidth]{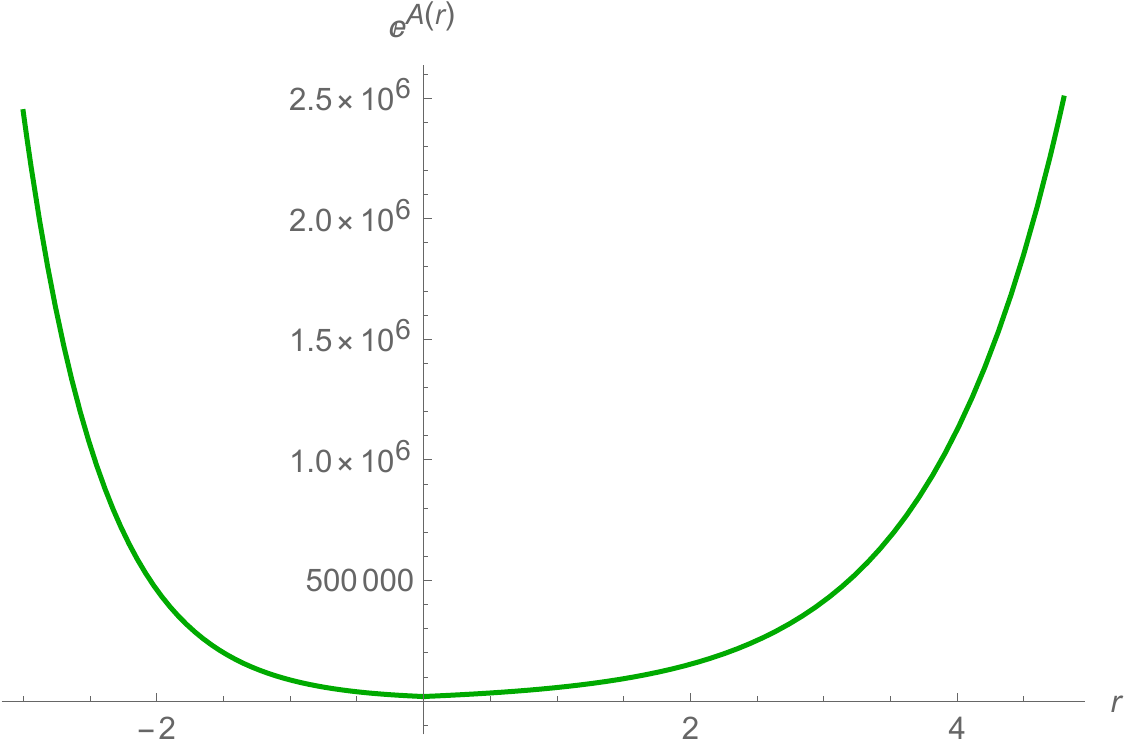}
                 \caption{Solution for $e^{A(r)}$}
         \end{subfigure}
         \caption{An example of RG-flow interfaces interpolating between $N=4$ $AdS_5$ vacuum on the right side and $N=2$ $AdS_5$ vacuum on the left side with $\ell=\kappa=1$, $h_1=-2$, $g_1=\sqrt{2}$ and $g_2=-\frac{\sqrt{2}}{4}$.}\label{fig4}
 \end{figure}

\begin{figure}
         \centering
               \begin{subfigure}[b]{0.35\textwidth}
                 \includegraphics[width=\textwidth]{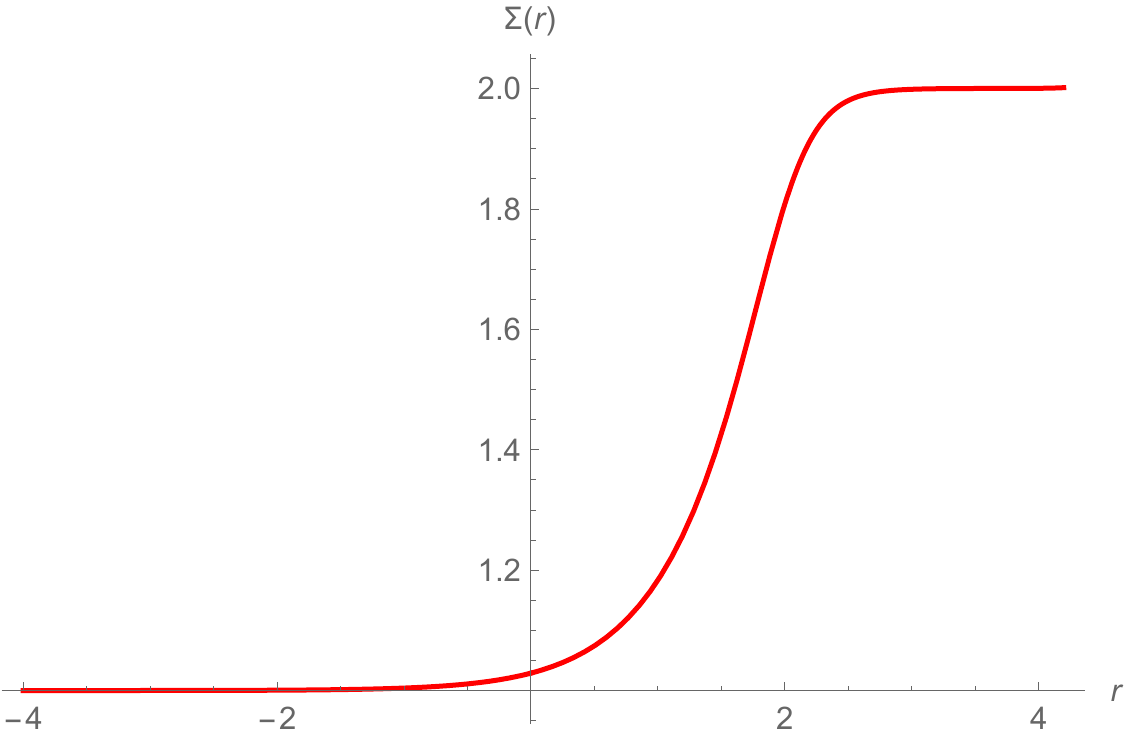}
                 \caption{Solution for $\Sigma(r)$}
         \end{subfigure}
         \begin{subfigure}[b]{0.35\textwidth}
                 \includegraphics[width=\textwidth]{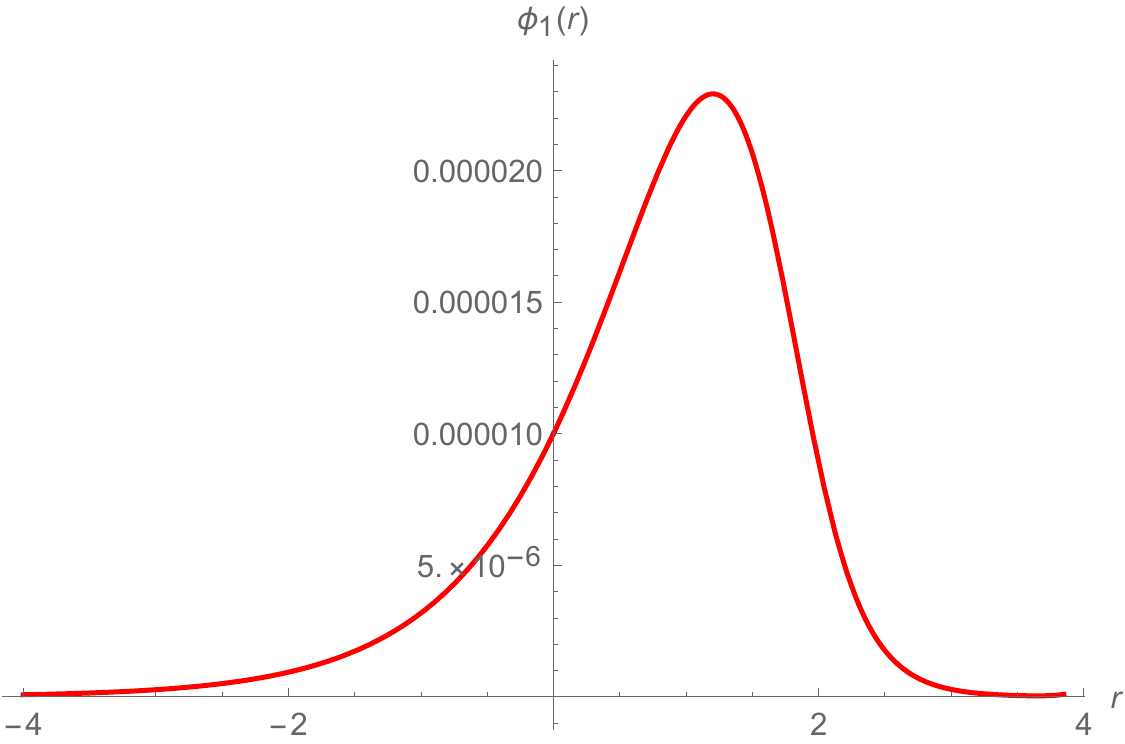}
                 \caption{Solution for $\phi_1(r)$}
         \end{subfigure}\\
         \begin{subfigure}[b]{0.35\textwidth}
                 \includegraphics[width=\textwidth]{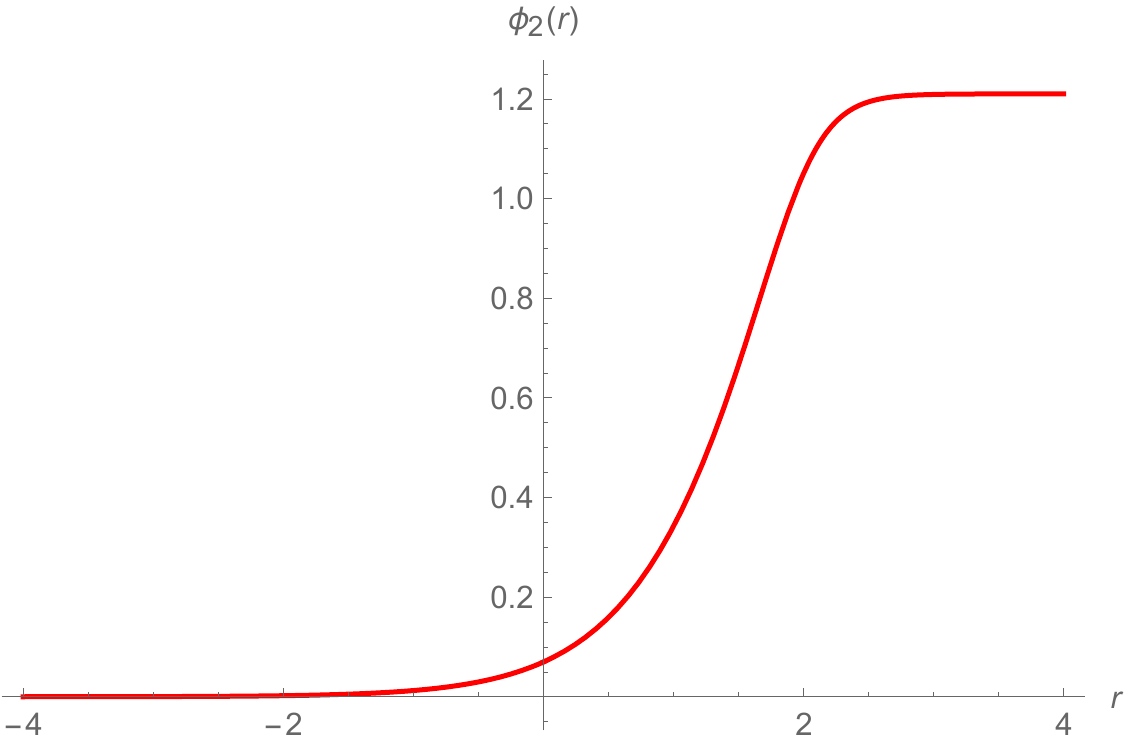}
                 \caption{Solution for $\phi_2(r)$}
         \end{subfigure}
         \begin{subfigure}[b]{0.35\textwidth}
                 \includegraphics[width=\textwidth]{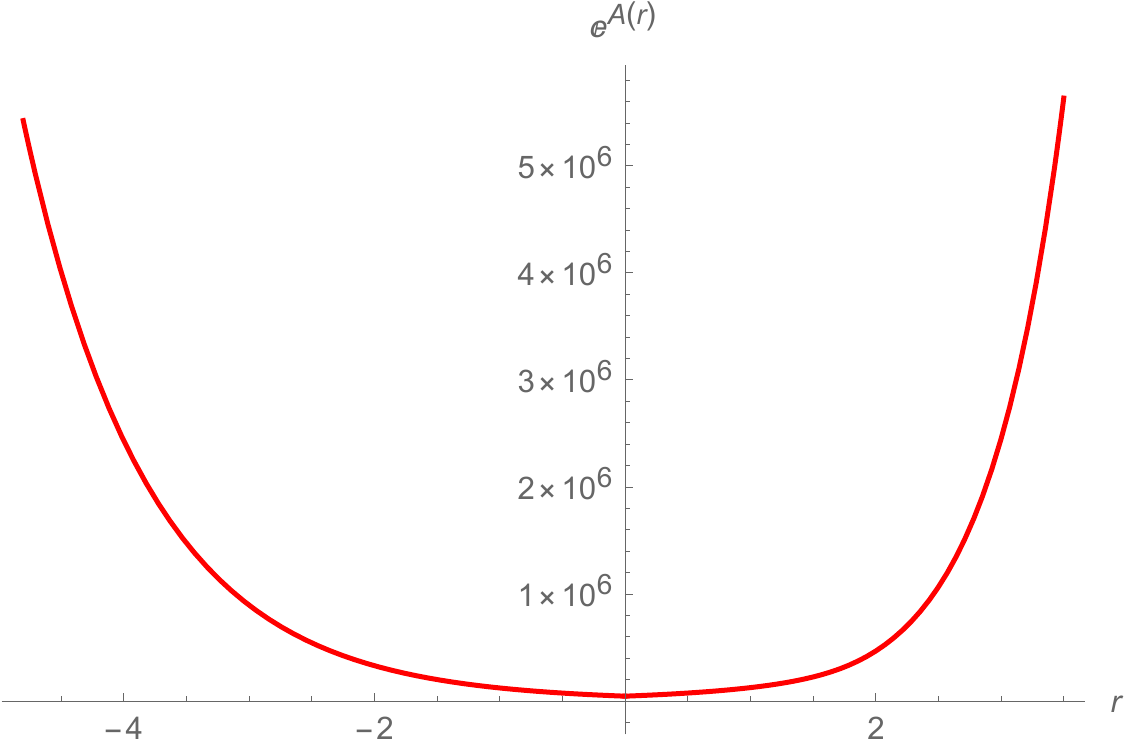}
                 \caption{Solution for $e^{A(r)}$}
         \end{subfigure}
         \caption{An example of RG-flow interfaces interpolating between $N=4$ $AdS_5$ vacuum on the left side and $N=2$ $AdS_5$ vacuum on the right side with $\ell=\kappa=1$, $h_1=-2$, $g_1=\sqrt{2}$ and $g_2=-\frac{\sqrt{2}}{4}$.}\label{fig5}
 \end{figure}

\begin{figure}
         \centering
               \begin{subfigure}[b]{0.45\textwidth}
                 \includegraphics[width=\textwidth]{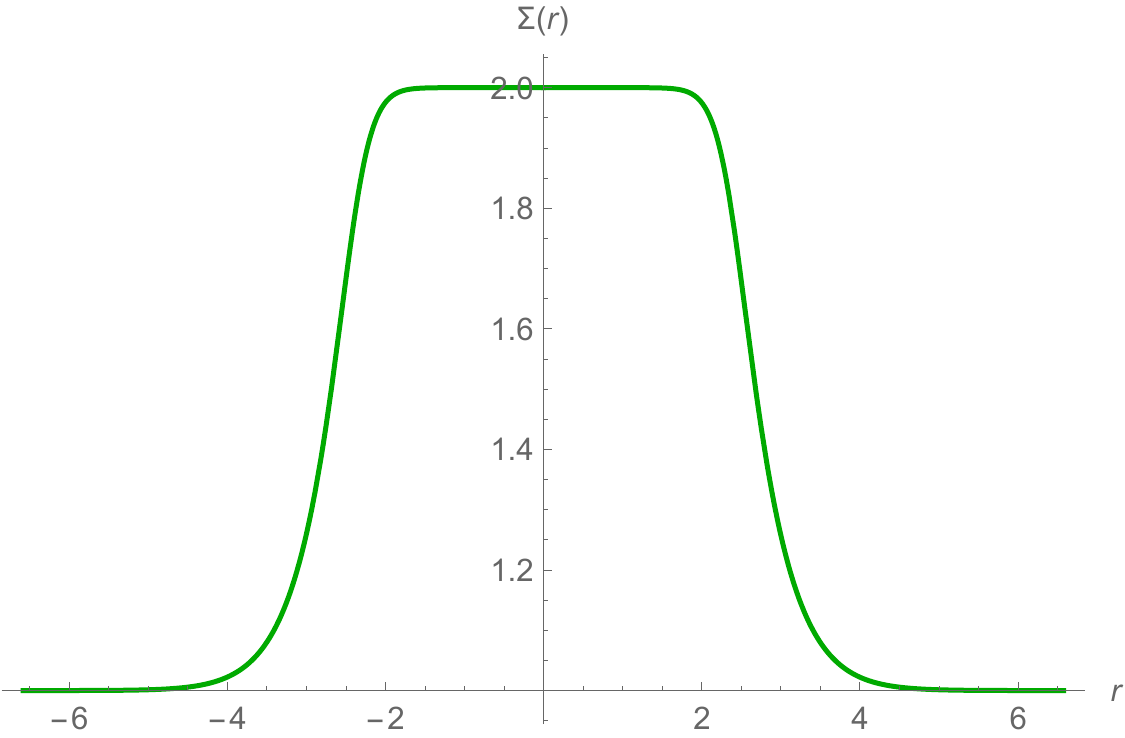}
                 \caption{Solution for $\Sigma(r)$}
         \end{subfigure}
         \begin{subfigure}[b]{0.45\textwidth}
                 \includegraphics[width=\textwidth]{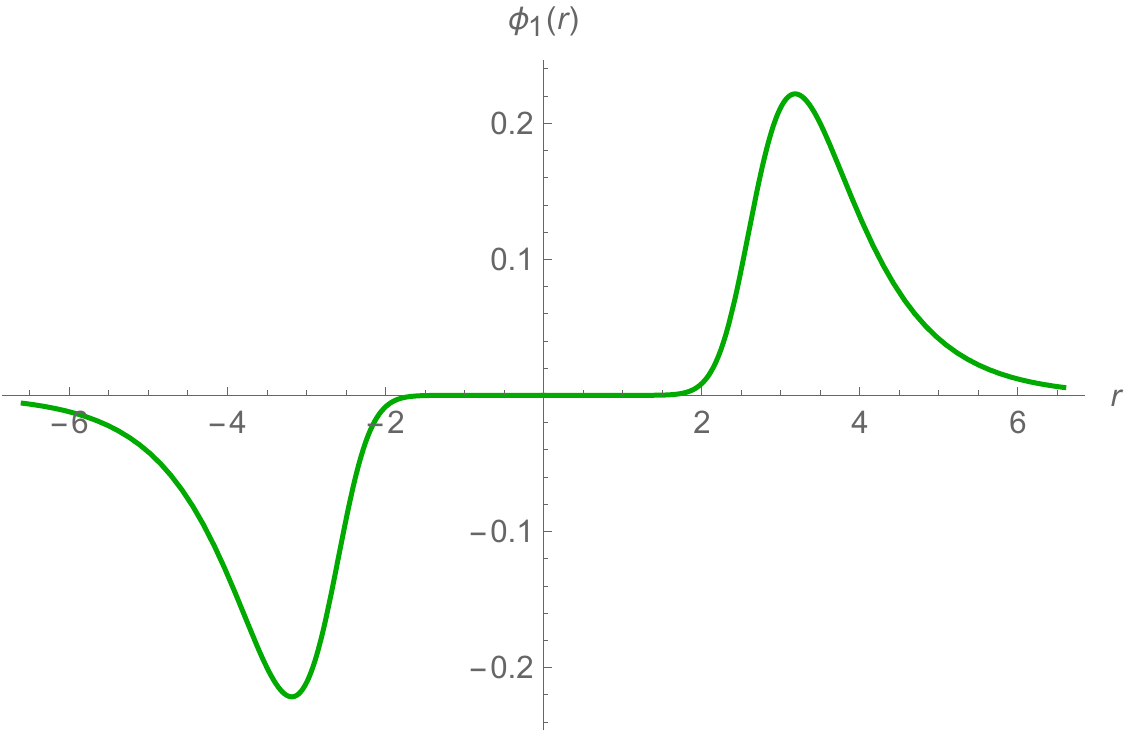}
                 \caption{Solution for $\phi_1(r)$}
         \end{subfigure}\\
         \begin{subfigure}[b]{0.45\textwidth}
                 \includegraphics[width=\textwidth]{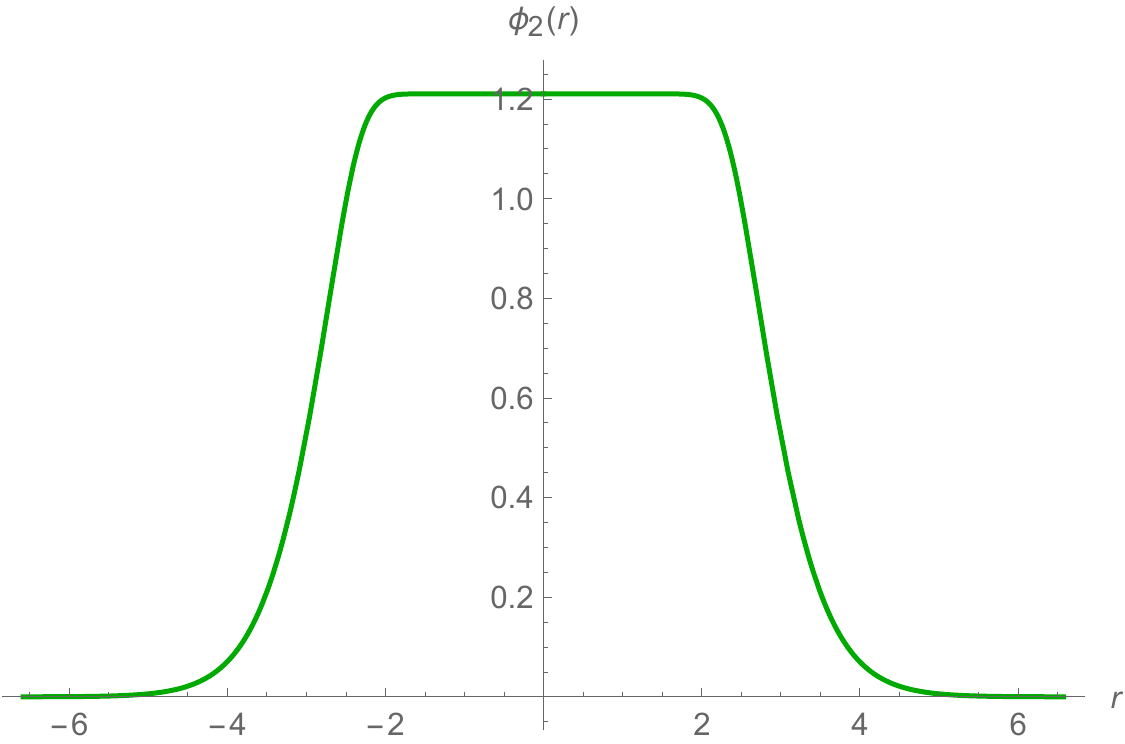}
                 \caption{Solution for $\phi_2(r)$}
         \end{subfigure}
         \begin{subfigure}[b]{0.45\textwidth}
                 \includegraphics[width=\textwidth]{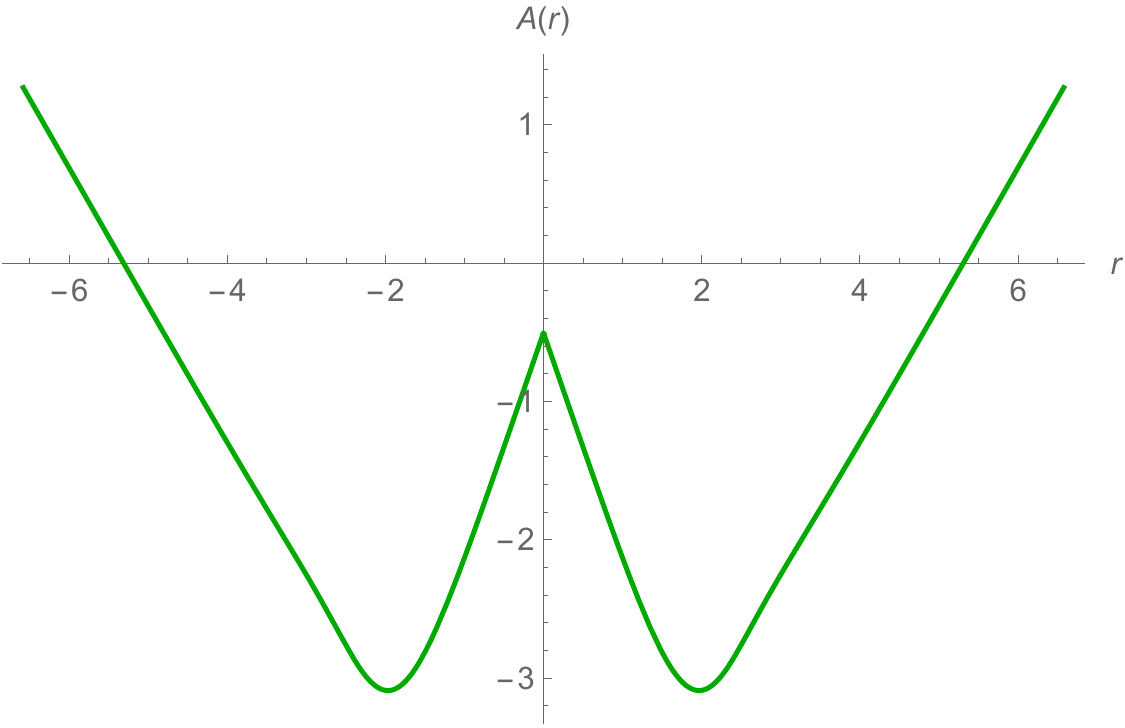}
                 \caption{Solution for $A(r)$}
         \end{subfigure}\\
         \begin{subfigure}[b]{0.45\textwidth}
                 \includegraphics[width=\textwidth]{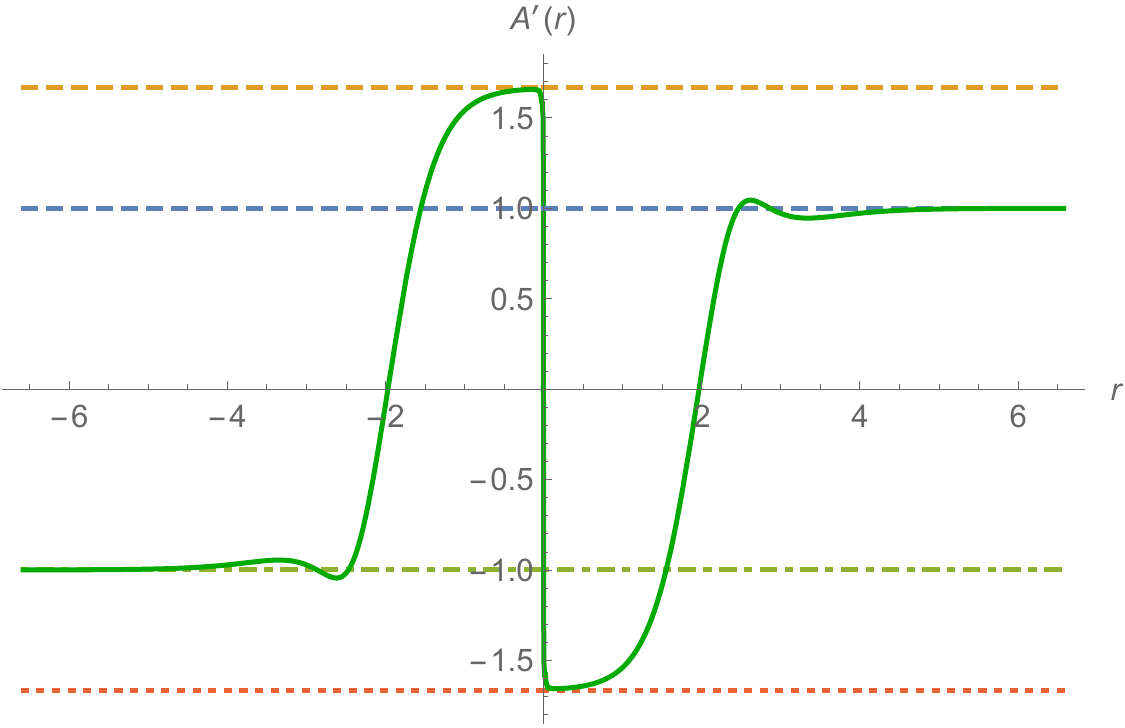}
                 \caption{Solution for $A'(r)$}
         \end{subfigure}
         \caption{An example of solutions with RG-flow interfaces on both sides of a Janus interface between $N=2$ $AdS_5$ vacua at the origin with $\ell=\kappa=1$, $h_1=-2$, $g_1=\sqrt{2}$ and $g_2=-\frac{\sqrt{2}}{4}$.}\label{fig6}
 \end{figure}

\begin{figure}
         \centering
               \begin{subfigure}[b]{0.45\textwidth}
                 \includegraphics[width=\textwidth]{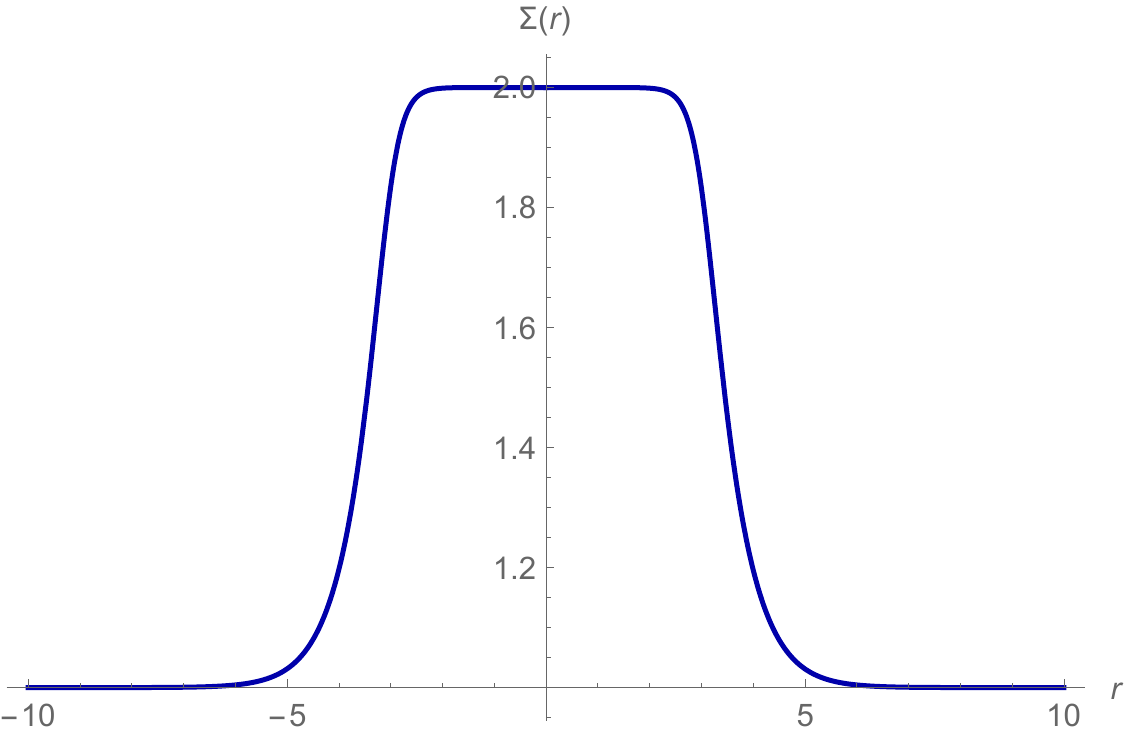}
                 \caption{Solution for $\Sigma(r)$}
         \end{subfigure}
         \begin{subfigure}[b]{0.45\textwidth}
                 \includegraphics[width=\textwidth]{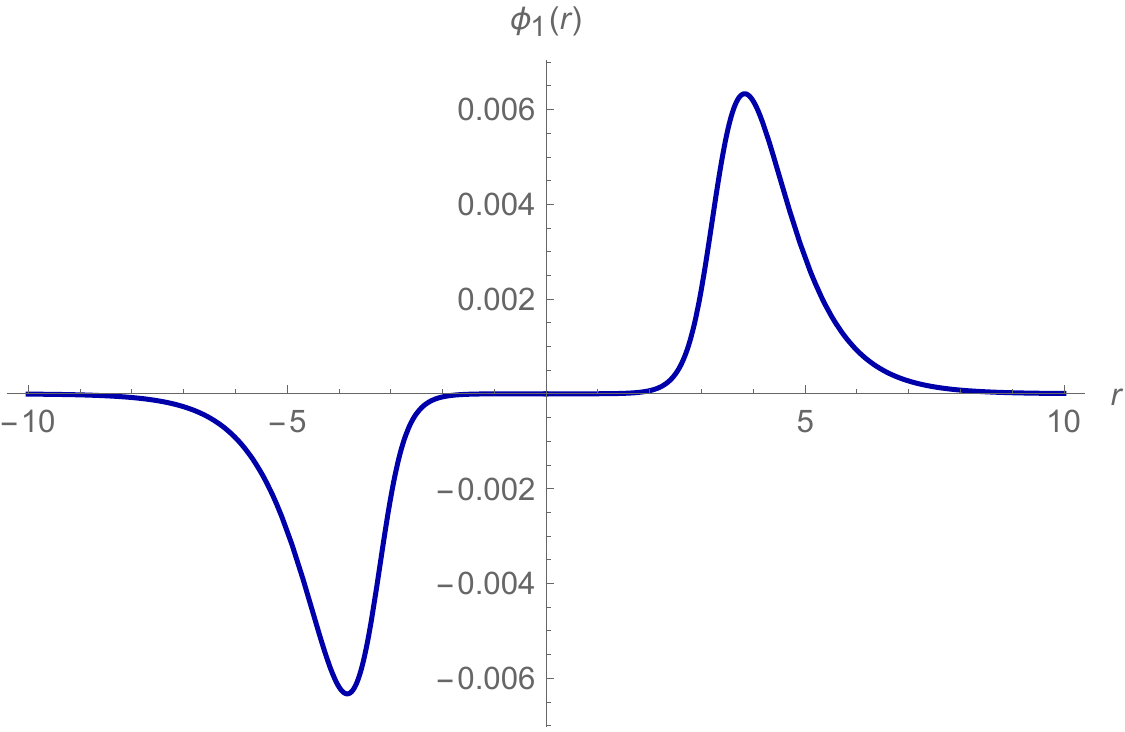}
                 \caption{Solution for $\phi_1(r)$}
         \end{subfigure}\\
         \begin{subfigure}[b]{0.45\textwidth}
                 \includegraphics[width=\textwidth]{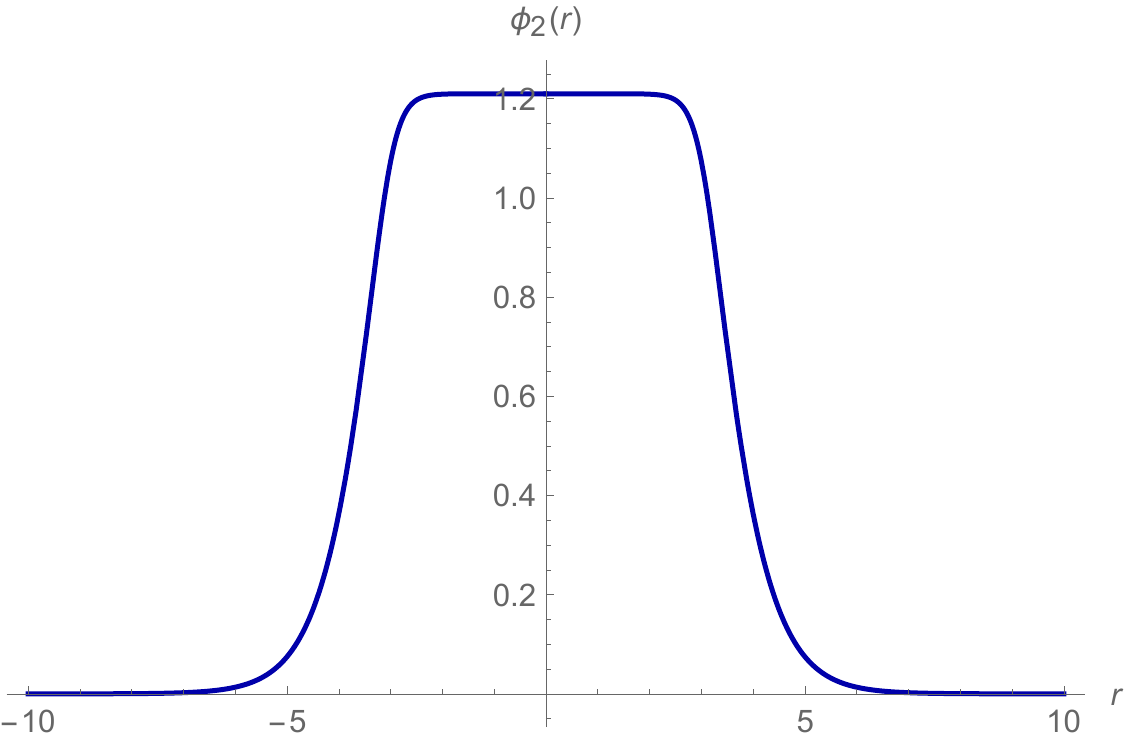}
                 \caption{Solution for $\phi_2(r)$}
         \end{subfigure}
         \begin{subfigure}[b]{0.45\textwidth}
                 \includegraphics[width=\textwidth]{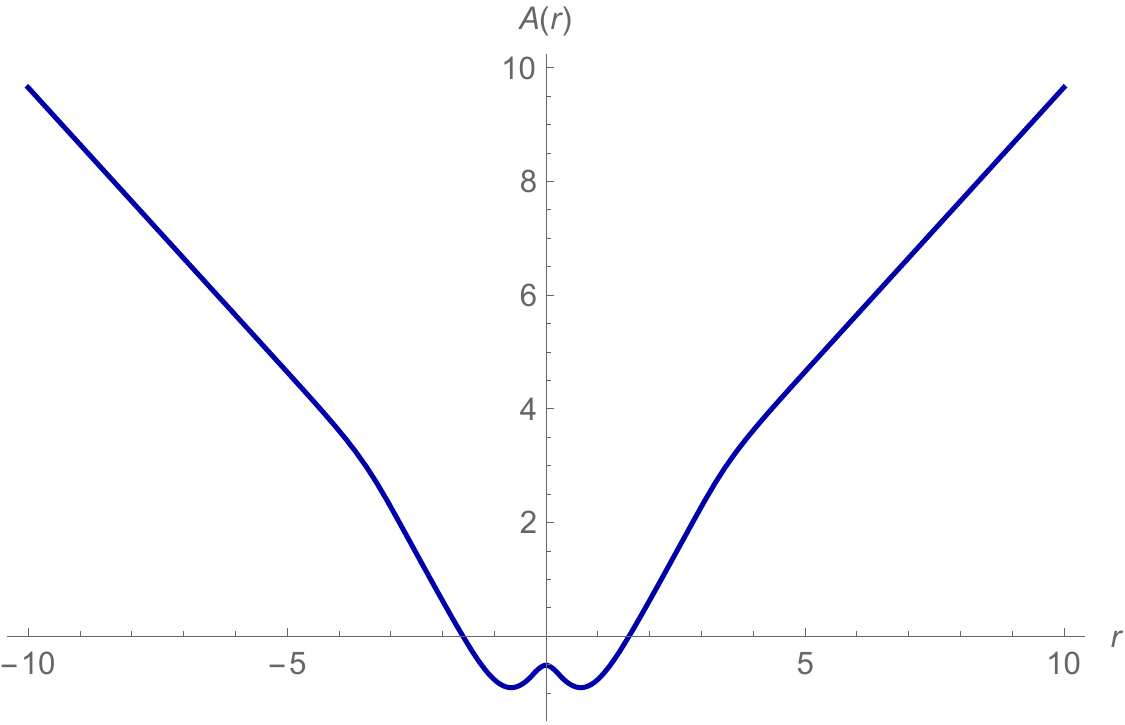}
                 \caption{Solution for $A(r)$}
         \end{subfigure}\\
         \begin{subfigure}[b]{0.45\textwidth}
                 \includegraphics[width=\textwidth]{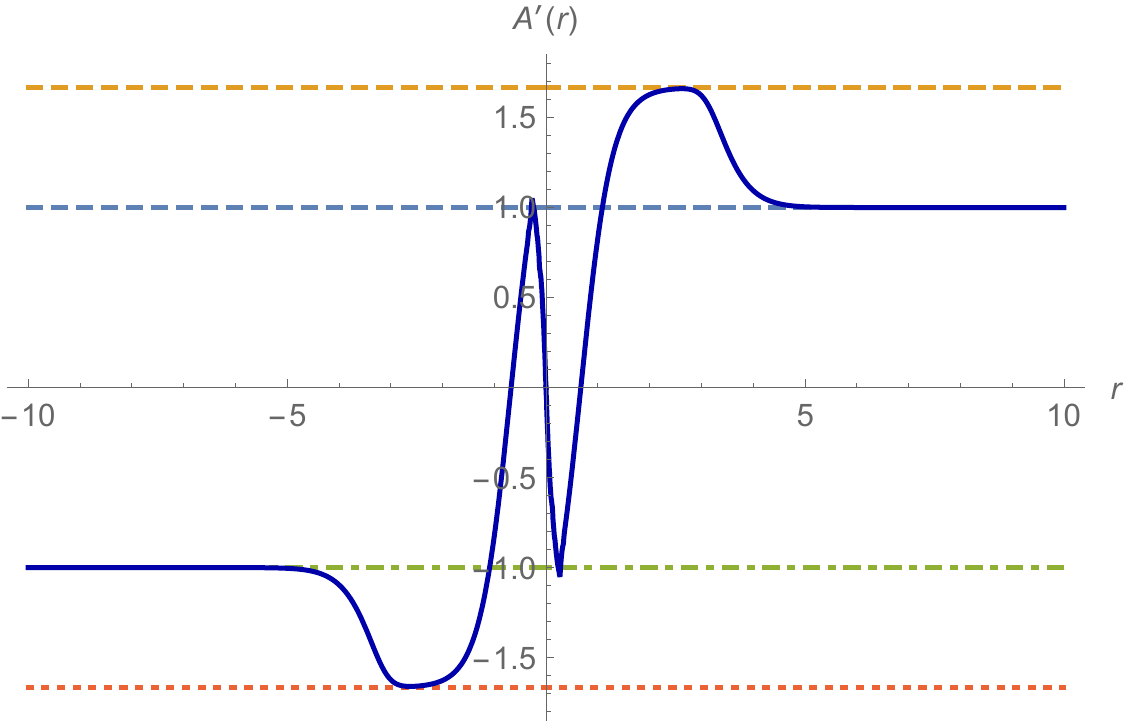}
                 \caption{Solution for $A'(r)$}
         \end{subfigure}
         \caption{An example of solutions with RG-flow interfaces on both sides of a Janus interface between $N=4$ $AdS_5$ vacua at the origin with $\ell=\kappa=1$, $h_1=-2$, $g_1=\sqrt{2}$ and $g_2=-\frac{\sqrt{2}}{4}$.}\label{fig7}
 \end{figure}

\begin{figure}
         \centering
               \begin{subfigure}[b]{0.35\textwidth}
                 \includegraphics[width=\textwidth]{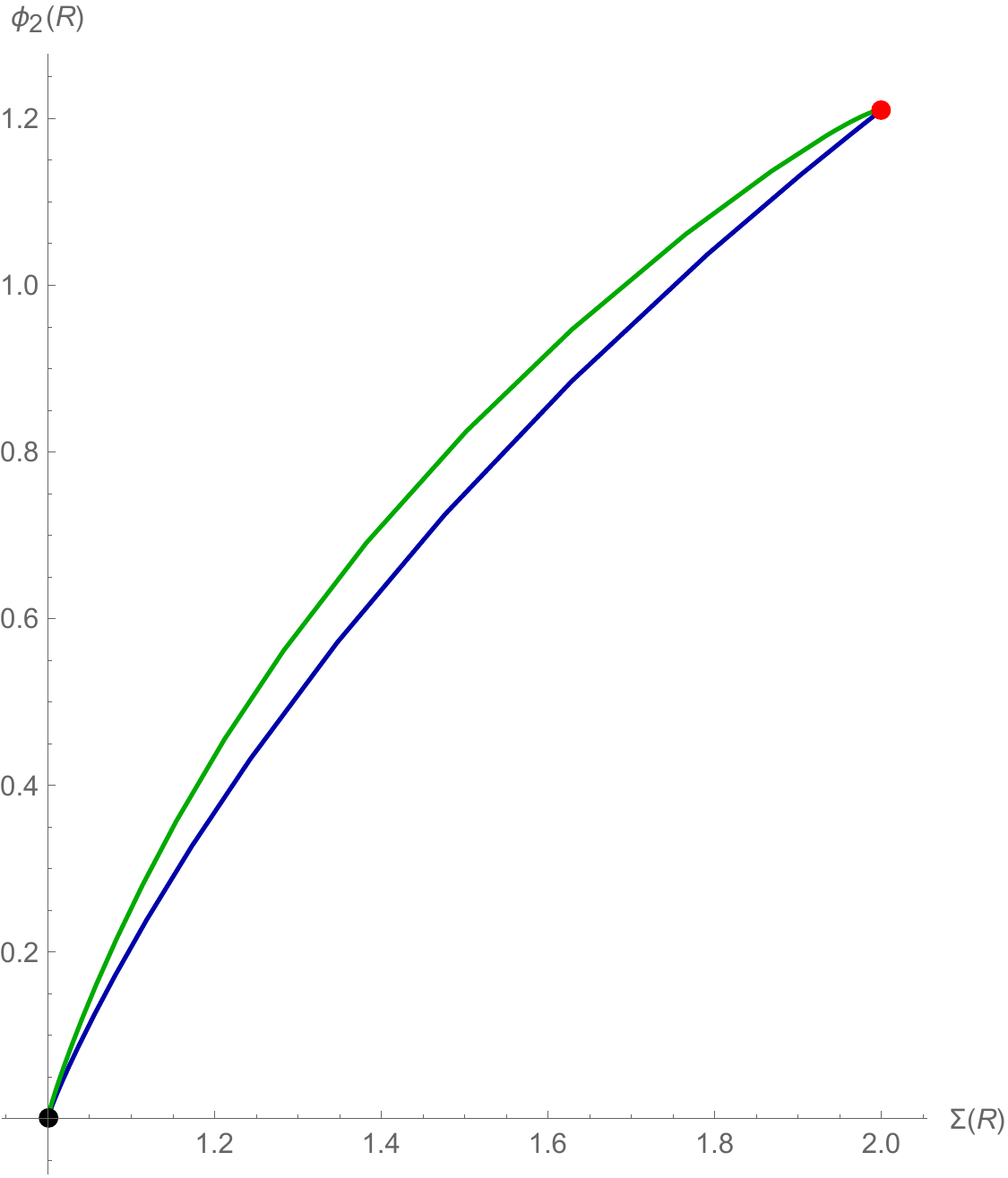}
                 \caption{Trajectories of the solutions on the left side of the interface at the origin}
         \end{subfigure}\hspace{1cm}
         \begin{subfigure}[b]{0.35\textwidth}
                 \includegraphics[width=\textwidth]{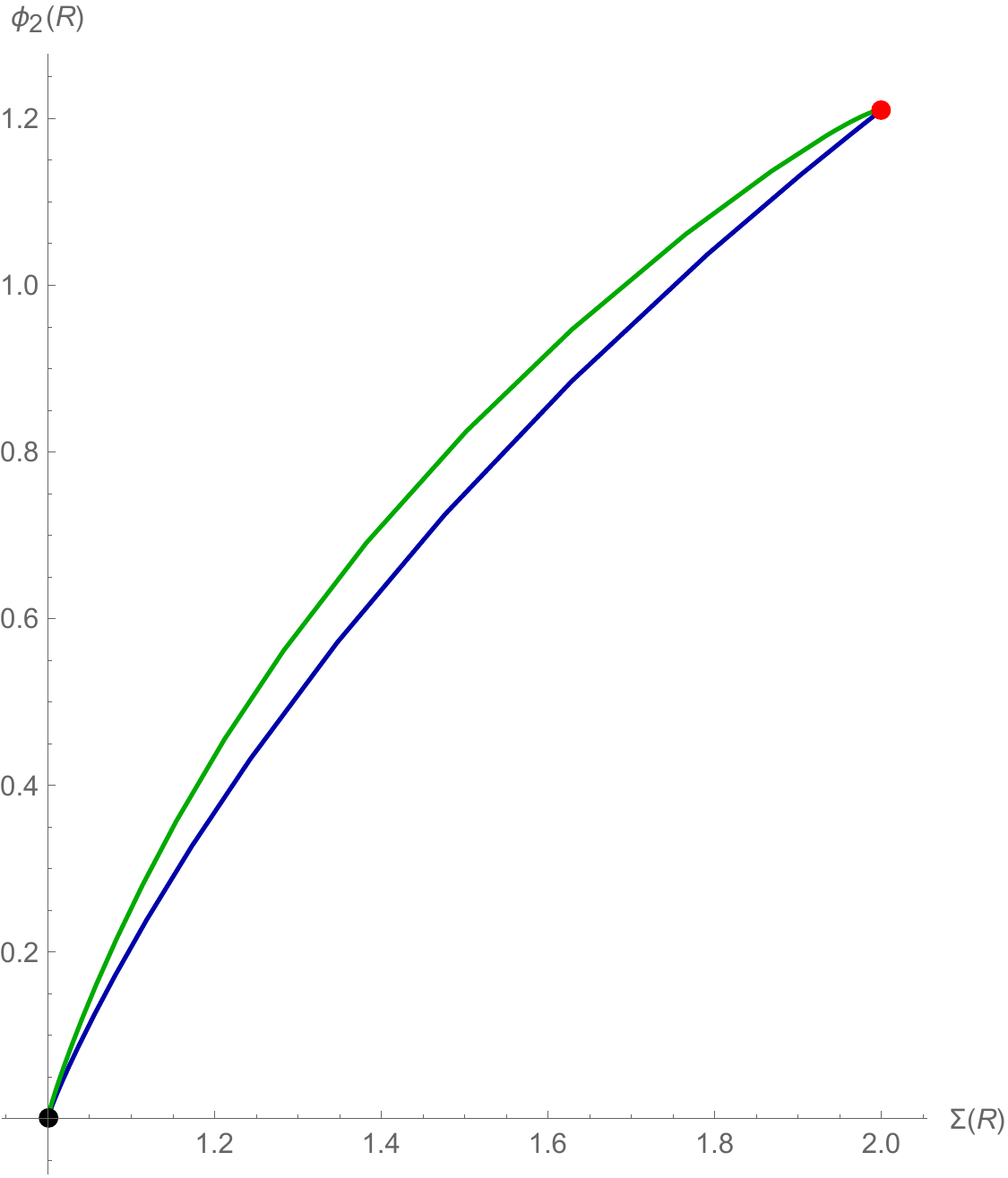}
                 \caption{Trajectories of the solutions on the right side of the interface at the origin}
         \end{subfigure}
         \caption{Trajectories of the solutions given in figures \ref{fig6} and \ref{fig7} on the $(\phi_2,\Sigma)$-plane with the black and red dots representing the $N=4$ and $N=2$ $AdS_5$ vacua, respectively.}\label{fig8}
 \end{figure}

\indent Further fine-tune the boundary conditions, we manage to obtain examples of RG-flow interfaces as shown in figures \ref{fig4} and \ref{fig5}. In this type of solutions, the two sides of the interfaces correspond to different SCFTs related by RG flows. A solution with the $N=4$ critical point on the right (left) and the $N=2$ critical point on the left (right) is shown in figure \ref{fig4} (\ref{fig5}). These solutions are similar to those found in six and three dimensions studied in \cite{6D_Janus,6D_Janus_RG} and \cite{3D_Janus2,3D_Janus3,3D_Janus4}. For $g_2\neq 0$, the asymptotic behaviors of scalars near the $N=4$ vacuum are given by
\begin{eqnarray}
& &\Sigma\sim e^{-\frac{2r}{L}},\qquad L=\frac{2}{g},\nonumber \\
& &\phi_1\sim e^{-\left(1+\frac{2}{\rho}\right)\frac{r}{L}},\qquad \phi_2\sim e^{-\left(2-\frac{2}{\rho}\right)\frac{r}{L}}\label{N4_g2_asymp}
\end{eqnarray}
For the sake of comparison with the results of \cite{5D_N4_flow_Davide}, we have redefined the coupling constants as follows
\begin{equation}
h_1=-g,\qquad g_1=\frac{g}{\sqrt{2}},\qquad g_2=-\frac{\sqrt{2}}{\rho}g\, .
\end{equation} 
We also note that the asymptotic expansion in \eqref{N4_g2_asymp} is valid only for $\rho\neq 2$ as pointed out in \cite{5D_N4_flow_Davide}. For $\rho=2$, a source term for a dimension-$2$ operator dual to $\Sigma$ also appears as in \eqref{N4_asymp}. In the numerical solutions, we have $g_2=-\frac{1}{4}g_1$ corresponding to $\rho=8$, so the deformation involves a vacuum expectation value of a dimension-$2$ operator dual to $\Sigma$ and source terms for operators of dimensions $\Delta_1=3-\frac{2}{\rho}=\frac{11}{4}$ and $\Delta_2=2+\frac{2}{\rho}=\frac{9}{4}$ dual to $\phi_1$ and $\phi_2$, respectively. It should also be noted that the dimensions of the dual operators are precisely in agreement with the scalar masses given in \cite{5D_flowII}. The asymptotic expansion near the $N=2$ $AdS_5$ critical point turns out to be much more complicated due to the mixing of all scalars. We refrain from presenting it here, but, according to the scalar masses given in \cite{5D_flowII}, the deformations should involve dual operators of dimensions
\begin{equation} 
\Delta_1=3+\sqrt{25-\frac{72}{2+\rho}},\quad \Delta_2=1+\sqrt{25-\frac{72}{2+\rho}},\quad \Delta_3=2+\sqrt{25-\frac{72}{2+\rho}}\, .
\end{equation}
\indent By extremely fine-tuning the boundary conditions to make the turning point very close to the $N=2$ $AdS_5$ critical point, we find solutions with different structures from all the previously found solutions. These are shown in figures \ref{fig6} and \ref{fig7}. In these solutions, there appear to be other turning points at which $A'=0$ apart from the chosen point at $r_0=0$. In figure \ref{fig6}, the solution takes the form of two RG-flow interfaces between $N=4$ and $N=2$ $AdS_5$ vacua joined at the origin, and the $N=2$ $AdS_5$ vacuum appears on both sides of the origin. By identifying the turning points as positions of the interfaces, it should be expected that there are three interfaces. Two of them at $r\approx \pm 2$ coorrespond to RG-flow interfaces between $N=1$ and $N=2$ SCFTs in four dimensions while the one at the origin $r=0$ describes a Janus interface between $N=1$ SCFTs on each side. 
\\
\indent A similar structure appears in the solution shown in figure \ref{fig7}. Although there are three turning points at which $A'=0$ as in the previous case, on each side of the origin, the $N=4$ $AdS_5$ vacuum appears rather than the $N=2$ vacuum. On the far left and far right sides, the $N=4$ phase undergoes an RG flow to the $N=2$ phase. The solution then proceeds through the turning points at $r\approx \pm 1$ to the $N=4$ $AdS_5$ critical point which appears on both sides of a Janus interface at the origin. The trajectories of the solutions on the $(\phi_2,\Sigma)$-plane given in figure \ref{fig8} also indicate that both sides of the solutions indeed interpolate between the two $AdS_5$ critical points. These solutions would be expected to give examples of multi-Janus interfaces.
 
\section{Conclusions and discussions}\label{conclusion}
We have studied supersymmetric Janus solutions in five-dimensional $N=4$ gauged supergravity with $SO(2)\times SO(3)$ and $SO(2)_D\times SO(3)$ gauge groups. In the former, there is only one supersymmetric $N=4$ $AdS_5$ vacuum, so there are only solutions describing conformal interfaces within the dual $N=2$ SCFT in four dimensions. In the latter, a number of more interesting solutions appears. Apart from solutions interpolating between the $N=4$ vacua, there are solutions in which the $N=4$ phases undergo holographic RG flows to the $N=2$ phases on both sides of the interfaces. By tuning the boundary conditions, the solutions can go very close to the $N=2$ vacuum and stay longer at this critical point. Following \cite{warner_Janus}, we can interpret these solutions as conformal interfaces between four-dimensional $N=1$ SCFTs dual to the $N=2$ $AdS_5$ critical points on each side of the interfaces. 
\\
\indent We  have also given examples of RG-flow interfaces between different SCFTs related to each other by RG flows on the two sides. Finally, the numerical analysis also leads to solutions with more than one turning points. These might naturally be interpreted as multi-interfaces in $N=1$ and $N=2$ SCFTs in four dimensions as in the solutions studied in \cite{multi_face_Janus} and \cite{5D_Janus_DHoker3}. Both of the $N=4$ gauged supergravities studied in this paper could be obtained from a consistent truncation of eleven-dimensional supergravity. Therefore, we also expect that the Janus solutions found here could lead to new supersymmetric Janus solutions in M-theory.       
\\
\indent As anticipated in \cite{5D_N4_flow_Davide}, upon uplifted to eleven dimensions, the $N=4$ $AdS_5$ vacuum should correspond to the $N=2$ Maldacena-Nunez SCFT obtained from M5-brane wrapped on a Riemann surface \cite{maldacena_nogo} while the $N=2$ $AdS_5$ vacuum would be identified with one of the $N=1$ SCFTs found in \cite{4D_SCFT_from_M5_Bobev}. Accordingly, we expect the Janus solutions obtained in this paper to describe conformal interfaces within these $N=2$ and $N=1$ SCFTs as well as RG-flow interfaces between the $N=1$ and $N=2$ SCFTs on each side of the interfaces. It would be interesting to precisely identify these conformal and RG-flow interfaces corresponding to various Janus solutions obtained in this work. As in the solutions found in \cite{RG-interface_Gauntlett}, these should involve position-dependent vacuum expectation values and turning on source terms for operators dual to scalar fields in the supergravity solutions. Working out an explicit truncation ansatz of eleven-dimensional supergravity on $M_3\times S^2\times S^1$ leading to the five-dimensional $N=4$ gauged supergravity considered here is also worth considering. The procedure involves solving a set of constraints given in \cite{Malek_AdS5_N4_embed}, and the result could be used to uplift the holographic RG flows given in \cite{5D_N4_flow_Davide} and Janus solutions in this paper to eleven dimensions. This would give a holographic description of RG flows and conformal interfaces within four-dimensional $N=1$ and $N=2$ SCFTs in the M-theory context.     


\begin{thebibliography}{99}
\bibitem{maldacena} J. M. Maldacena, ``The large $N$ limit of superconformal field theories and supergravity'', Adv. Theor. Math. Phys. \textbf{2} (1998) 231-252, arXiv: hep-th/9711200.
\bibitem{Gubser_AdS_CFT} S. S. Gubser, I. R. Klebanov and A. M. Polyakov, ``Gauge Theory Correlators from Non-Critical String Theory'', Phys. Lett. \textbf{B428} (1998) 105-114, arXiv: hep-th/9802.109.
\bibitem{Witten_AdS_CFT} E. Witten, ``Anti De Sitter Space and holography'', Adv. Theor. Math. Phys. \textbf{2} (1998) 253-291, arXiv: 9802150.
\bibitem{Bak_Janus} D. Bak, M. Gutperle and S. Hirano, ``A Dilatonic Deformation of $AdS_5$ and its Field Theory Dual'', JHEP 05 (2003) \textbf{072}, arXiv: hep-th/0304129.
\bibitem{5D_Janus_CK} A. Clark and A. Karch, ``Super Janus'', JHEP 10 (2005) \textbf{094}, arXiv: hep-th/0506265.
\bibitem{Freedman_Janus} A. B. Clark, D. Z. Freedman, A. Karch and M. Schnabl, ``Dual of the Janus solution: An interface conformal field theory'', Phys. Rev. \textbf{D71} (2005)
066003, arXiv: hep-th/0407073.
\bibitem{DHoker_Janus} E. D' Hoker, J. Estes and M. Gutperle, ``Interface Yang-Mills, supersymmetry, and Janus'', Nucl. Phys. \textbf{B753} (2006) 16, arXiv: hep-th/0603013.
\bibitem{Witten_Janus} D. Gaiotto and E. Witten, ``Janus Configurations, Chern-Simons Couplings, And The thetaAngle in N=4 Super Yang-Mills Theory'', JHEP 1006 (2010) 097, arXiv: 0804.2907.
\bibitem{Freedman_Holographic_dCFT} O. DeWolfe, D. Z. Freedman and H. Ooguri, ``Holography and Defect Conformal Field Theories'', Phys. Rev. \textbf{D66} (2002) 025009, arXiv: hep-th/0111135.
\bibitem{6D_Janus} M. Gutperle, J. Kaidi and H. Raj, ``Janus solutions in six-dimensional gauged supergravity'', JHEP 12 (2017) \textbf{018}, arXiv: 1709.09204.
\bibitem{6D_Janus_RG} P. Karndumri, ``Janus and RG-flow interfaces from matter-coupled $F(4)$ gauged supergravity'', Phys. Rev. \textbf{D111} (2025) 026013, arXiv: 2405.17169.
\bibitem{ISO3_flow} P. Karndumri, ``Holographic RG flows and Janus interfaces from ISO(3)$\times$U(1) $F(4)$ gauged supergravity'', Phys. Rev. \textbf{D111} (2025) 026022, arXiv: 2409.20151.
\bibitem{5D_Janus_Suh} M. W. Suh, ``Supersymmetric Janus solutions in five and ten dimensions'', JHEP 09 (2011)
\textbf{064}, arXiv: 1107.2796.
\bibitem{Bobev_5D_Janus1} N. Bobev, F. F. Gautason, K. Pilch, M. Suh, J. van Muiden, ``Janus and J-fold Solutions from Sasaki-Einstein Manifolds'', Phys. Rev. \textbf{D100} (2019) 081901, arXiv: 1907.11132.
\bibitem{Bobev_5D_Janus2} N. Bobev, F. F. Gautason, K. Pilch, M. Suh, J. van Muiden, ``Holographic Interfaces in $N=4$ SYM: Janus and J-folds'', JHEP 05 (2020) \textbf{134}, arXiv: 2003.09154.
\bibitem{warner_Janus} N. Bobev, K. Pilchand N. P. Warner, ``Supersymmetric Janus Solutions in Four Dimensions'', JHEP 1406 (2014) 058, arXiv: 1311.4883.
\bibitem{N3_Janus} P. Karndumri, ``Supersymmetric Janus solutions in four-dimensional $N=3$ gauged supergravity'', Phys. Rev. \textbf{D93} (2016) 125012, arXiv: 1604.06007.
\bibitem{N3_JanusII} P. Karndumri, ``New Janus interfaces from four-dimensional $N=3$ gauged supergravity'', Eur. Ohys. J. \textbf{C84} (2024) 1059, arXiv: 2408.00424.
\bibitem{tri-sasakian-flow} P. Karndumri, ``Supersymmetric deformations of 3D SCFTs from tri-sasakian truncation'', Eur. Phys. J. C (2017) \textbf{77}, 130, arXiv: 1610.07983.
\bibitem{orbifold_flow} P. Karndumri and K. Upathambhakul, ``Supersymmetric RG flows and Janus from type II orbifold compactification'', Eur. Phys. J. C (2017) \textbf{77}, 455, arXiv: 1704.00538.
\bibitem{Minwoo_4DN8_Janus} M. Suh, ``Supersymmetric Janus solutions of dyonic $ISO(7)$-gauged $N=8$ supergravity'', JHEP 04 (2018) \textbf{109}, arXiv: 1803.00041.
\bibitem{Kim_Janus} N. Kim and S. J. Kim, ``Re-visiting Supersymmetric Janus Solutions: A Perturbative Construction'', Chin. Phys. \textbf{C44} (2020) 7, 073104, arXiv: 2001.06789.
\bibitem{N5_flow} P. Karndumri and C. Maneerat, ``Supersymmetric solutions from $N=5$ gauged supergravity'', Phys. Rev. \textbf{D101} (2020) 126015, arXiv: 2003.05889.
\bibitem{N6_flow} P. Karndumri and J. Seeyangnok, ``Supersymmetric solutions from $N=6$ gauged supergravity'', Phys. Rev. \textbf{D103} (2021) 066023, arXiv: 2012.10978.
\bibitem{3D_Janus_de_Boer} C. Bachas, J. de Boer, R. Dijkgraaf, and H. Ooguri, ``Permeable conformal walls and
holography'', JHEP 06 (2002) \textbf{027}, arXiv:hep-th/0111210.
\bibitem{3D_Janus_Bachas} C. Bachas and M. Petropoulos, ``Anti-de-Sitter D-branes'', JHEP 02 (2001) \textbf{025}, arXiv:hep-th/0012234.
\bibitem{3D_Janus_Bak} D. Bak, M. Gutperle and S. Hirano, ``Three dimensional Janus and time-dependent black
holes'', JHEP 02 (2007) \textbf{068}, arXiv: hep-th/0701108.
\bibitem{half_BPS_AdS3_S3_ICFT} M. Chiodaroli, M. Gutperle and D.
Krym, ``Half-BPS Solutions locally asymptotic to $AdS_3\times S^3$
and interface conformal field theories'', JHEP 02 (2010)
\textbf{066}, arXiv: 0910.0466.
\bibitem{exact_half_BPS_string} M. Chiodaroli, E. D'Hoker, Y, Guo
and M. Gutperle, ``Exact half-BPS string-junction solutions in
six-dimensional supergravity'', JHEP 12 (2011) \textbf{086}, arXiv:
1107.1722.
\bibitem{multi_face_Janus} D. Bak and H. Min, ``Multi-faced Black Janus and Entanglement'', JHEP 03 (2014) \textbf{046}, arXiv: 1311.5259.
\bibitem{3D_Janus} K. Chen and M. Gutperle ``Janus solutions in three-dimensional N=8 gauged supergravity'', JHEP 05 (2021) \textbf{008}, arXiv: 2011.10154.
\bibitem{N8_omega_Janus} P. Karndumri and C. Maneerat, ``Supersymmetric Janus solutions in $\omega$-deformed $N=8$ gauged supergravity'', Eur. Phys. J. \textbf{C81} (2021) 801, arXiv: 2012.15763.
\bibitem{N4_Janus} P. Karndumri, ``Holographic RG flows and Janus solutions from matter-coupled $N=4$ gauged supergravity'', Eur. Phys. J. \textbf{C81} (2021) 520, arXiv: 2102.05532.
\bibitem{ISO7_Janus} P. Karndumri and C. Maneerat, ``Janus solutions from dyonic $ISO(7)$ maximal gauged supergravity'', JHEP 10 (2021) \textbf{117}, arXiv: 2108.13398.
\bibitem{3D_Janus2} K. Chen, M. Gutperle and C. Hultgreen-Mena, ``Janus and RG-flow interfaces in three-dimensional gauged supergravity'', JHEP 03 (2022) \textbf{057}, arXiv: 2111.01839.
\bibitem{N4_omega_Janus} T. Assawasowan and P. Karndumri, ``New supersymmetric Janus solutions from $N=4$ gauged supergravity'', Phys. Rev. \textbf{D105} (2022) 106004, arXiv: 2203.03413.
\bibitem{3D_Janus3} M. Gutperle and C. Hultgreen-Mena, ``Janus and RG interfaces in three-dimensional gauged supergravity II: General $\alpha$'', JHEP 08 (2022) \textbf{126}, arXiv: 2205.10398.
\bibitem{3D_Janus4} M. Gutperle and C. Hultgreen-Mena, ``Janus and RG-interfaces in minimal 3d gauged supergravity'', arXiv: 2412.16749.
\bibitem{Guarino_Janus_M} A. Anabalon, M. Chamorro-Burgos and A. Guarino, ``Janus and Hades in M-theory'', JHEP 11 (2022) \textbf{150}, arXiv: 2207.09287.
\bibitem{Guarino_S_fold_Janus} A. Guarino and M. Suh, ``S-folds and $AdS_3$ flows from the D3-brane'', JHEP 11 (2022) \textbf{134}, arXiv: 2207.14015.
\bibitem{4D_Janus_from_11D} E. D'Hoker, J. Estes, M. Gutperle and D. Krym, ``Janus solutions in M-theory'', JHEP
06 (2009) \textbf{018}, arXiv: 0904.3313.
\bibitem{5D_Janus_DHoker1} E. D'Hoker, J. Estes and M. Gutperle, ``Ten-dimensional supersymmetric Janus solutions'', Nucl. Phys. \textbf{B757} (2006) 79, arXiv: hep-th/0603012.
\bibitem{5D_Janus_DHoker2} E. D'Hoker, J. Estes and M. Gutperle, ``Exact half-BPS Type IIB interface solutions. I. Local solution and supersymmetric Janus'', JHEP 06 (2007) \textbf{021}, arXiv: 0705.0022.
\bibitem{5D_Janus_DHoker3} E. D'Hoker, J. Estes and M. Gutperle, ``Exact half-BPS Type IIB interface solutions. II: Flux solutions and multi-Janus'', JHEP 06 (2007) \textbf{022}, arXiv: 0705.0024.
\bibitem{RG-interface_Gauntlett} I. Arav, K. C. Matthew Cheung, J. P. Gauntlett, M. M. Roberts and C. Rosen, ``Superconformal RG interfaces in holography'', JHEP 11 (2020) \textbf{168}, arXiv: 2007.07891.
\bibitem{5D_N4_flow_Davide} N. Bobev, D. Cassani and H. Triendl, ``Holographic RG Flows for Four-dimensional $N=2$ SCFTs'', JHEP 06(2018)\textbf{086}, arXiv: 1804.03276.
\bibitem{5D_flowII} P. Karndumri, ``$AdS_5$ vacua and holographic RG flows from 5D $N=4$ gauged supergravity'', Eur. Phys. J. \textbf{C83} (2023) 164, arXiv: 2209.05270.
\bibitem{Malek_AdS5_N4_embed} E. Malek and V. V. Camell, ``Consistent truncations around half-maximal AdS5 vacua of 11-dimensional supergravity'', Class. Quant. Grav. \textbf{39} (2022) 7, 075026, arXiv: 2012.15601.
\bibitem{5D_N4_black_stringII} P. Karndumri, ``New supersymmetric $AdS_5$ black strings from 5D $N=4$ gauged supergravity'', Eur. Phys. J. \textbf{C83} (2023) 432, arXiv: 2211.07456.
\bibitem{N4_gauged_SUGRA} J. Schon and M. Weidner, ``Gauged $N=4$ supergravities'', JHEP 05 (2006) \textbf{034}, arXiv: hep-th/0602024.
\bibitem{5D_N4_Dallagata} G. Dall’Agata, C. Herrmann, and M. Zagermann, ``General matter coupled $N=4$ gauged
supergravity in five-dimensions'', Nucl. Phys. \textbf{B612} (2001) 123–150, arXiv: hep-th/0103106.
\bibitem{AdS5_N4_Jan} J. Louis, H. Triendl and M. Zagermann, ``$N = 4$ supersymmetric $AdS_5$ vacua and their moduli spaces'', JHEP 10 (2015) \textbf{083}, arXiv:1507.01623.
\bibitem{N2_holography_Bobev} N. Bobev, H. Elvang, D. Z. Freedman and S. S. Pufu, ``Holography for $N=2^*$ on $S^4$'', JHEP 07 (2014) \textbf{001}, arXiv: 1311.1508.
\bibitem{mass_deform_5D_SCFT} M. Gutperle, J. Kaidi and H. Raj, ``Mass deformations of 5d SCFTs via holography'', JHEP 02 (2018) \textbf{165}, arXiv: 1801.00730.
\bibitem{5D_N4_curved_DW} M. Zagermann, ``$N=4$ ``Fake'' Supergravity'', Phys.Rev. \textbf{D71} (2005) 125007, arXiv: hep-th/0412081.
\bibitem{maldacena_nogo} J. Maldacena and C. Nunez, ``Supergravity description of field theories on curved manifolds and a no go theorem'', Int. J. Mod. Phys. \textbf{A16} (2001) 822, arXiv: hep-th/0007018.
\bibitem{4D_SCFT_from_M5_Bobev} I. Bah, C. Beem, N. Bobev and B. Wecht, ``Four-dimensional SCFTs from M5-branes'', JHEP 06 (2012) \textbf{005}, arXiv: 1203.0303.
\end{thebibliography}
\end{document}